%% file: main.tex
\documentclass{article}

    \usepackage[preprint]{neurips_2024}

\usepackage[utf8]{inputenc} 
\usepackage[T1]{fontenc}    
\usepackage{hyperref}       
\usepackage{url}            
\usepackage{booktabs}       
\usepackage{amsfonts,amsthm,amsmath,amssymb}       
\usepackage{nicefrac}       
\usepackage{microtype}      
\usepackage{xcolor}         

\usepackage{tikz}
\usetikzlibrary{shapes,arrows,positioning,quotes}

\usepackage{algorithm}
\usepackage{algorithmic}

\usepackage{longtable,tcolorbox,pifont,booktabs}
\usepackage{bm, dsfont, amsmath, amsfonts, amssymb}

\newtheorem{theorem}{Theorem}
\newtheorem{lemma}{Lemma}
\newtheorem*{lemma*}{Lemma}

\newtheorem{corollary}{Corollary}
\newtheorem{assumption}{Assumption}

\newtheorem{definition}{Definition}
\newtheorem{remark}{Remark}

\newcommand{\bR}{\mathbb{R}}
\newcommand{\bN}{\mathbb{N}}

\newcommand{\bP}{\mathbb{P}}

\newcommand{\cS}{\mathcal{S}}
\newcommand{\cR}{\mathcal{R}}
\newcommand{\cA}{\mathcal{A}} 
\newcommand{\br}{\bm{R}} 
\newcommand{\cH}{\mathcal{H}} 
\newcommand{\cX}{\mathcal{X}} 

\newcommand{\cD}{\mathcal{D}} 

\newcommand{\nag}{\operatorname{NG}} 
\DeclareMathOperator{\ts}{Ts}

\DeclareMathOperator*{\argmax}{arg\,max}

\newcommand{\eqdef}{:=}
\newcommand\numberthis{\addtocounter{equation}{1}\tag{\theequation}}

\newcommand{\ie}{\textit{i.e.} }

\newcommand{\cf}{\textit{cf.} }
\newcommand{\algo}{\texttt{TBRVI}~}
\usepackage{mathrsfs}
\DeclareRobustCommand{\bigO}{\mathcal{O}}

\title{Learning in Zero-Sum Markov Games: Relaxing Strong Reachability and Mixing Time Assumptions}

\author{%
  Reda Ouhamma \quad Maryam Kamgarpour \\
  SYCAMORE Lab, \'Ecole Polytechnique F\'ed\'erale de Lausanne (EPFL),\\
  1015 Lausanne, Switzerland.
}

\begin{document}

\maketitle

\begin{abstract}
  We address payoff-based decentralized learning in infinite-horizon zero-sum Markov games. In this setting, each player makes decisions based solely on received rewards, without observing the opponent's strategy or actions nor sharing information. Prior works established finite-time convergence to an approximate Nash equilibrium under strong reachability and mixing time assumptions. We propose a convergent algorithm that significantly relaxes these assumptions, requiring only the existence of a single policy (not necessarily known) with bounded reachability and mixing time.  Our key technical novelty is introducing Tsallis entropy regularization to smooth the best-response policy updates. By suitably tuning this regularization, we ensure sufficient exploration, thus bypassing previous stringent assumptions on the MDP. By establishing novel properties of the value and policy updates induced by the Tsallis entropy regularizer, we prove finite-time convergence to an approximate Nash equilibrium.
\end{abstract}

\section{Introduction}
Markov games are a class of multi-agent decision-making problems with a rich history dating back to the foundational work of \citet{shapley1953stochastic}. In Markov games, a popular solution concept is a Nash equilibrium, a set of strategies such that no player can improve her payoff by unilaterally changing her strategy. Computing a Nash equilibrium for general Markov games is computationally intractable \citep{daskalakis2009complexity, chen2009settling}. However, in the special case of zero-sum Markov games, where the players have opposing interests, Nash equilibria can be computed efficiently given known dynamics and rewards. Specifically, \citet{shapley1953stochastic} showed that a stationary Markovian Nash equilibrium can be efficiently computed in discounted zero-sum Markov games. This tractability has led to extensive approaches on computing equilibria in zero-sum Markov games, such as variations of policy and value iterations \citep{shapley1953stochastic, hoffman1966nonterminating, pollatschek1969algorithms, van1978discounted, filar1991algorithm}.
Recently, learning in zero-sum Markov games has gained increasing attention, where the goal is learning a Nash equilibrium given only observations of rewards and transitions, without knowing the environment \citep{littman1994markov}. The two key objectives in this setting are \citep{bowling2001rational}: rationality, which requires players to converge to their opponent's best response if the opponent's strategy is asymptotically stationary; and convergence, which refers to reaching a Nash equilibrium. 

A particularly desirable framework for learning in Markov games is payoff-based decentralized learning, where players learn without coordinating their actions or observing their opponent's rewards, strategy, and actions. This setting is attractive due to its scalability and lower communication requirements. However, \cite{liu2022learning} proved that learning a Nash equilibrium in this framework is intractable without further assumptions. To address this, recent works have adopted self-play, where both players use the same learning algorithm. Self-play has been instrumental in solving hard games such as GO, StarCraft, and Dota 2 \citep{silver2017mastering,vinyals2019grandmaster,Berner2019Dota2W}, and has led to theoretical advances in zero-sum Markov games. Existing self-play algorithms with provable convergence to a Nash equilibrium primarily fall into two categories: policy gradient methods, which rely on direct optimization of the strategy \citep{wei2021last, daskalakis2020independent,zhao2022provably,cen2021fast}; and value-based methods, which rely on value function estimates to guide strategy updates, including $Q$-learning-type algorithms \citep{sayin2021decentralized} and value iteration methods \citep{chen2023finite}. 


In this paper, we study infinite horizon discounted zero-sum Markov games. We aim to provide a finite-time convergence for payoff-based decentralized learning without relying on past restrictive assumptions, namely, strong reachability conditions and uniformly bounded mixing times. In the following, we discuss these limitations in more detail and introduce our approach, which relaxes these assumptions while maintaining strong convergence guarantees.

\paragraph{Reachability assumption.} A popular assumption on Markov decision processes (MDPs) is strong reachability \citep{wei2017online,wei2021last, chen2021sample,cai2024uncoupled}. Namely, there exists an integer $L$ such that \emph{for any strategies} used by players, the expected time to visit any state from any state is bounded by $L$. Under this assumption, many algorithms were proposed and demonstrated either average-iterate convergence to an $\epsilon$-approximate Nash equilibrium \citep{wei2021last} or last-iterate convergence \citep{cai2024uncoupled}. Past efforts that relax strong reachability have weak convergence guarantees. For example, \citet{chen2023finite, sayin2021decentralized} demonstrate convergence up to a bias due to entropy regularization. Specifically, \cite{sayin2021decentralized} provides an algorithm with an asymptotic best-iterate convergence to a biased Nash equilibrium. The previous result was extended in \citet{chen2023finite} to a finite-time convergence to a biased Nash equilibrium. 

Note that the reachability assumption is tightly connected with exploration in the MDP. The consideration of an episodic setting (finite horizon) \citep{daskalakis2020independent} or access to a generative model \citep{zhao2022provably} bypasses the reachability challenges. In the former case, in every new episode, the state is reset and one can avoid being stuck in a state from which reachability is challenged. In the latter, one can query any state and hence, exploration is not an issue. 


\paragraph{Mixing-time assumption.} This common assumption requires a uniform upper bound on the speed of convergence of Markov chains induced by any strategy to their stationary distributions \citep{chen2021sample}. The necessity of this assumption comes from the need to estimate value functions from time-inhomogeneous payoffs, which are due to players changing their strategies with time. This challenge is also relevant for single agent actor-critic algorithms \citep{xu2020improved,xu2020improving,kumar2023sample,olshevsky2023small}. For zero-sum Markov games, the algorithm proposed in \citet{chen2023finite} also relaxes the above mixing-time assumption, though it only shows convergence to a biased Nash equilibrium as explained above.

The requirement on bounding mixing times arises due to the payoff-based setting as the players need to estimate value functions, and because mixing times bound the speed at which the value functions converge. Note that this requirement would not be needed if players could coordinate to fix their policies when querying the MDP \citep{wei2021last, chen2021sample}.

Building on this discussion of the objectives and past works in the zero-sum Markov games, we formulate the aim of this paper through the following question: 


\begin{center}
\textbf{\textit{In zero-sum Markov games, can we learn an approximate Nash equilibrium efficiently without assumptions of strong reachability and uniform mixing times?}}
\end{center}



\paragraph{Contributions.} We address payoff-based decentralized zero-sum Markov games, focusing on the fundamental problem of learning an approximate Nash equilibrium, weakening the above two assumptions. In particular, we only assume the existence of a single strategy that induces an irreducible Markov chain with a finite mixing time, significantly weakening strong reachability and uniform mixing-time assumptions. Our contributions include
\begin{itemize}
    \item A decentralized, convergent, and rational algorithm;
    \item A finite-time sample complexity for learning an approximate Nash equilibrium.
\end{itemize}
Our key contribution is using Tsallis-entropy regularization for zero-sum Markov games. Originally derived from statistical physics, Tsallis entropy generalizes Shannon entropy \citep{tsallis1988possible} and has gained increasing attention in the online learning community for its effectiveness in both stochastic and adversarial settings \citep{zimmert2021tsallis}. This work introduces Tsallis entropy within the smoothed best-response algorithm of \cite{chen2023finite}, demonstrating its distinct advantages in Markov games due to its improved exploration properties. Specifically, we establish:
\begin{itemize}
    \item  Lower bounds on the strategies of our algorithm, ensuring sufficient exploration (see Lemma \ref{lem:margins}) and polynomial upper bounds on the mixing times; 
    \item The smoothness and strong convexity of a Tsallis-entropy regularized operator, enabling the convergence of our strategies (see Lemma \ref{lem:lipschitz_tsallis}).
\end{itemize}
\textbf{Organization.} In Section 2, we introduce the problem setting, establish the notation, and discuss the limiting assumptions of past work. Section 3 presents our proposed algorithm and highlights our contributions, including a theorem with the convergence rate to an approximate Nash equilibrium, along with corollaries of finite-time sample complexity and rationality. Finally, Section 4 provides a proof sketch outlining the high-level arguments and emphasizing our key technical contributions.

\section{Preliminaries}


\paragraph{Notations.} We denote the sets of real and natural numbers by $\bR$ and $\bN$, respectively. The opponent of player $i \in \{1,2\}$ is denoted by $-i$. The probability simplex over a finite space $\cX$ is $\Delta^\cX$.

\subsection{Setting}

\paragraph{Setting.} We study infinite-horizon two-player zero-sum Markov games, defined as a tuple $\mathcal{M} = (\cS, \cA^1, \cA^2, P, \cR^1, \cR^2, \gamma)$. $\cS$ is a finite state space and $\cA^1$ and $\cA^2$ are finite action spaces for players 1 and 2, respectively, we denote $A_{\max} = \max(|\cA^1|, |\cA^2|)$. The transition kernel $P: \cS\times\cA^1\times\cA^2 \mapsto \Delta_{\cS}$ specifies the probability $P(s' |s, a^1, a^2)$ of transition from $s$ to $s'$ with actions $a^1$ and $a^2$. The reward function is denoted as $\cR: \cS \times\cA^1 \times\cA^2 \mapsto \bR$ for player 1 and  $-\cR$ for player 2, with $\max_{s,a^1,a^2}|\cR(s,a^1,a^2)| \le 1$ for all $s,a^1,a^2$. The discount factor denoted by $\gamma$ satisfies $0< \gamma < 1$. We consider a fixed initial state $s_0$, without loss of generality, see \citet{fiechter1994efficient}. 

\paragraph{Payoff-based information.} We consider a payoff-based decentralized information setting, where players follow the same decision-making algorithm (self-play). In this setup, each player makes decisions based only on her obtained rewards. Moreover, players do not know the other player's strategy or actions and do not share information. At time $k$, the players observe the state $s_k$, choose actions $(a_k^i)_{i={1,2}}$, and the state transitions to $s_{k+1}$. Subsequently, the players observe their respective rewards $\cR(s_k,a_k^i,a_k^{-i})$ and $- \cR(s_k,a_k^i,a_k^{-i})$, and update their strategies independently. 




A stationary strategy for player $i \in \{1,2\}$ is a mapping $\pi^i$ from $S$ to $\Delta^{\cA^i}$. We denote the joint strategy $(\pi^1, \pi^2)$ as $\pi$. We now define the $q$-functions:
\begin{align*}
    q_\pi^1\left(s, a^1\right) &= \mathbb{E}_\pi \left[\sum_{t=0}^{\infty} \gamma^t \mathcal{R}\left(s_t, a_t^1, a_t^{2}\right) \mid \!s_0\!=\!s, a_0^1 = a^1\right],\\
    q_\pi^2\left(s, a^2\right) &= -\mathbb{E}_\pi \left[\sum_{t=0}^{\infty} \gamma^t \mathcal{R}\left(s_t, a_t^1, a_t^{2}\right) \mid \!s_0\!=\!s, a_0^2 = a^2\right],
\end{align*}
where the expectation is over the randomness of the strategy $\pi$ and of the transitions. Finally, the value function of players of player $i$, $i=1,2$ as $v_\pi^i(s) \eqdef \mathbb{E}_{a^i \sim \pi^i(\cdot \mid s)}\left[q_\pi^i\left(s, a^i\right)\right]$ 

\begin{definition}{(Nash equilibrium)}
    The strategies $(\pi_{\text{NE}}^1, \pi_{\text{NE}}^2)$ are a Nash equilibrium (NE) if for $i \in \{1,2\}$, $s\in \cS$, and any $\pi^i \in (\Delta^{\cA_i})^\cS$:
    \begin{equation*}
        v_{\pi_{\text{NE}}^{i}, \pi_{\text{NE}}^{-i}}^i(s) \ge v_{\pi^{i}, \pi_{\text{NE}}^{-i}}^i(s).
    \end{equation*}
\end{definition}
The above means that no agent can improve her value by unilaterally changing her strategy.

\begin{definition}{(Approximate Nash equilibrium)} 
\label{def:Approximate_NE}
    Given a strategy $(\pi^i,\pi^{-i})$, we define the Nash Gap:
    \begin{equation*}
        \nag(\pi^i,\pi^{-i}) = \max_{s\in \cS} \left\{\sum_{i=1,2}  \left( \max_{\hat{\pi}^i \in \Delta^{\cA^i}}v_{\hat{\pi}^i,\pi^{-i}}^i(s) \!- \!v_{(\pi^i,\pi^{-i})}^i(s) \right)\right \},
    \end{equation*}
    For an $\epsilon>0$, the strategies $(\pi^i,\pi^{-i})$ are an $\epsilon$-approximate Nash equilibrium if $\nag(\pi^i,\pi^{-i}) \le \epsilon$.    
\end{definition}
The above definition shows that the Nash Gap is zero if and only if the strategies constitute a Nash equilibrium. Our objective in this paper is to learn an approximate Nash equilibrium in polynomial time without relying on assumptions of strong reachability or uniformly bounded mixing times.

Note that other notions of convergence were proposed in previous work, such as average-iterate convergence and no-regret guarantees. An average-iterate result entails a bound on the average Nash gap over strategies played throughout the iterations of the algorithm. On the other hand, a no-regret guarantee would involve each player's regret being sublinear given the other player's sequence of played strategies throughout the iterates of the algorithm. In normal-form games, a no-regret guarantee implies an average-iterate convergence to a Nash equilibrium \citep{freund1999adaptive}. However, no-regret does not imply convergence of the iterates to an approximate Nash equilibrium \citep{muthukumar2020impossibility}. Thus, the notion of convergence we are after is stronger than no-regret and average-iterate convergence.

\subsection{Limiting assumptions in state-of-the-art}
\label{sec:limiting_asspts}


\paragraph{Strong reachability.} We begin with the common strong reachability assumption \citep{auer2006logarithmic,chen2021sample}, also known as the irreducible game assumption in \cite{cai2024uncoupled}. 
\begin{definition}{(Strong reachability)}\label{as:strong_reachability}
    The MDP satisfies strong reachability if there exists a finite constant $L>0$ such that:
    \begin{equation*}
        \max_{s,s' \in \cS} \max_{\pi \in (\Delta^{\cA_i})^\cS \times (\Delta^{\cA_{-i}})^\cS} T_{s \mapsto s'}^\pi \le L,
    \end{equation*}
    where $T_{s \mapsto s'}^\pi$ is the expected time to reach state $s'$ from state $s$ when players follow strategy $\pi$.
\end{definition}
In particular, the above definition means that strong reachability implies that the Markov chain induced by any strategy is irreducible, namely, any two states are reachable from each other by a sequence of transitions with positive probability. Hence, this is an assumption on how easy it is to explore in an MDP. It was shown in \citet[Theorem 5.5.11]{durrett2019probability} that an irreducible Markov chain induced by a strategy $\pi$ with stationary distribution $\mu_\pi$ satisfies
\begin{equation}\label{eq:Equivalent_Reachability}
    T_{s \mapsto s}^\pi = 1/ \mu_\pi (s).
\end{equation}
Therefore, if the Markov chain induced by any strategy $\pi$ is irreducible, then strong reachability is equivalent to the stationary distribution being uniformly lower bounded $\mu_\pi (s) \ge 1/L$ for all states and strategies. The assumption of a positive lower bound (uniform with respect to all strategies) on the stationary distribution had been prevalent in reinforcement learning \citep{agarwal2021theory,mei2020global,zhang2022gradient}. However, it is very restrictive. To illustrate, consider the left MDP provided in Figure \ref{fig:MDP_exp}, and consider the strategy $\pi$ parameterized by $\xi \in [0,1]$ defined as:
\begin{equation}\label{eq:exp_policy}
    \pi(0,a) = \xi \text{ and } \quad \pi(0,b) = 1-\xi.
\end{equation}
Then, the corresponding stationary distribution $\mu_\pi$ is given by:
\begin{equation}\label{eq:exp_stationary}
    \mu_\pi = \left(\frac{1}{2+\xi}, \frac{1}{2}, \frac{\xi}{4+2\xi}\right),
\end{equation}
which implies $\lim_{\xi\to 0} \mu_\pi(2) = 0$. This invalidates Assumption \ref{as:strong_reachability} because there couldn't exist a positive $L$ such that $\min_s \inf_\pi \mu_\pi (s) \ge 1/L$, see Equation \eqref{eq:Equivalent_Reachability}. 

\begin{figure}[!t]
    \parbox{1.8in}{%
        \begin{tikzpicture}[-> , >= stealth',shorten >=1pt, line width=0.4pt ,
    node distance=2.2cm, inner sep=.2pt]
            \node[circle, draw , minimum size=.4cm](zero){0};
            \node[circle , draw , minimum size=.4cm](one)[right of= zero]{1};
            \node[circle , draw , minimum size=.4cm](two)[right of= one]{2};
            \path (zero) edge [ loop left ] node {$b\!: 1/2$} (zero) ;
            \path(zero) edge [bend left=2] node [above] {$b\!: 1/2$} (one) ;
            \path(zero) edge [bend left = 35] node [ above ] {$a\!: 1/2$} (one) ;
            \path(zero) edge [bend left = 56] node [ above ] {$a\!: 1/2$} (two) ;
            \path(one) edge [bend left=5] node [ below ] {$1/2$}(zero) ;
            \path (one) edge [ loop above ] node {$1/2$} (one) ;
            \path(two) edge [bend left=0] node [ below ] {$1/2$}(one) ;
            \path(two) edge [bend left=26] node [ below ] {$1/2$}(zero) ;
        \end{tikzpicture}
    }%
    \qquad\qquad\qquad
    \begin{minipage}{1.2in}%
        \begin{tikzpicture}[-> , >= stealth',shorten >=1pt, line width=0.4pt ,
    node distance=2.4cm, inner sep=.2pt]
            \node[circle, draw , minimum size=.4cm](zero){0};
            \node[circle , draw , minimum size=.4cm](one)[right of= zero]{1};
            \node[circle , draw , minimum size=.4cm](two)[right of= one]{2};
            \path (zero) edge [ loop left ] node {$b\!: 1/2$} (zero) ;
            \path(zero) edge [bend left=-6] node [above] {$b\!: 1/2$} (one) ;
            \path(zero) edge [bend left = 25] node [ above ] {$a, d\!: 1$} (one) ;
            \path(zero) edge [bend left = 55] node [ above ] {$\:\:a, c: 1/2$} (one) ;
            \path(zero) edge [bend left = 67] node [ above ] {$a, c\!: 1/2$} (two) ;
            \path(one) edge [bend left=15] node [ below ] {$1/2$}(zero) ;
            \path (one) edge [ loop above ] node {$1/2$} (one) ;
            \path(two) edge [bend left=0] node [ below ] {$1/2$}(one) ;
            \path(two) edge [bend left=30] node [ below ] {$1/2$}(zero) ;
        \end{tikzpicture}
    \end{minipage}%
    \caption{Two MDPs with three states: the transitions from states $1$ and $2$ are action independent. The arrows indicate the possible transitions labeled with their probabilities. \textbf{The left} is a single agent MDP with two actions $a$ and $b$ in state $0$. \textbf{The right} is a two-player MDP where the second player has two actions $c$ and $d$ in state $0$.}
    \label{fig:MDP_exp}
    \vspace*{-.3cm}
\end{figure}

In single-agent reinforcement learning, \citet{auer2008near} proposed a weaker alternative assumption. Namely, \citet{auer2008near} proved minimax optimal bounds requiring only that, for each pair of states, there exists a strategy under which the expected time to travel between them is bounded: $\max_{s,s'} \min_{\pi} T_{s \mapsto s'}^\pi \le L$. In multi-agent settings, however, relaxing strong reachability is more challenging because players can block each other from reaching certain states. For example, in the MDP on the right side of Figure \ref{fig:MDP_exp}, the second player can effectively prevent player 1 from reaching state 2 by consistently choosing action $d$. This highlights that a game setting exacerbates reachability challenges compared to a single-agent setting. To our knowledge, all previous works proving convergence to a Nash equilibrium in zero-sum Markov games have required strong reachability.  



\paragraph{Uniform mixing-times.} This assumption asserts the existence of a uniform upper-bound, with respect to the strategies, on the mixing times of the Markov chain induced by any strategy \citep{olshevsky2023small,chen2024finite,bhandari2018finite,wu2020finite}.


To state this assumption, we introduce two necessary notations. For a strategy $\pi=(\pi_1,\pi_2)$ and a transition matrix $P$, we define $P_\pi$ as the transition kernel induced by strategy $\pi$, and $P_\pi^k$ as the $k$-step transition kernel for an integer $k\in\bN$. 

\begin{definition}\label{def:MixingTime}
The $\epsilon$-mixing time of strategy $\pi$, with unique stationary distribution $\mu_\pi$, is defined as: $ t_{\pi, \epsilon}\!=\!\min \!\left\{k \geq 0\!:\! \max _{s \in \mathcal{S}}\left\|P_\pi^k(\cdot | s) \!-\! \mu_\pi(\cdot)\right\|_{\mathrm{TV}} \leq \epsilon\right\}$, where $\|.\|_{\mathrm{TV}}$ is the total variation distance. 
\end{definition}


\begin{definition}{(Assumption of uniform mixing-time)}\label{as:bounded_mixing}
    The MDP satisfies the uniform mixing-time assumption if for any $\epsilon>0$, there exists a finite mixing-time $t_{mix}(\epsilon)$ such that 
    \begin{equation*}
        \forall \pi \in (\Delta^{\cA_i})^\cS \times (\Delta^{\cA_{-i}})^\cS, \quad t_{\pi, \epsilon} \le t_{mix}(\epsilon).
    \end{equation*}
\end{definition}

Assuming a finite bound on the supremum of the mixing times over strategies as in Definition \ref{as:bounded_mixing} is very restrictive. To illustrate, we show that for the left MDP in Figure \ref{fig:MDP_exp} with the strategies defined in Equation \ref{eq:exp_policy} it holds that the mixing time grows to infinity when $\xi$ tends to one (see Appendix \ref{app:counteexample}). This invalidates the assumption in Definition \ref{as:bounded_mixing} because the mixing times are not uniformly bounded.\\ 
In single-agent reinforcement learning, the assumption of uniform mixing time can be relaxed for algorithms that perform many value function updates per strategy update, see \cite{kumar2023sample}. Intuitively, this assumption characterizes how fast value function estimates achieve their steady-state value. While this approach is acceptable in a single-agent setting, applying it in a multi-agent context would require significant coordination between the players to synchronize their updates.

\section{Algorithm and sample complexity}


\subsection{Algorithm}

We present our algorithm, Tsallis smoothed Best-Response Dynamics with Value Iteration, \algo. It builds on the algorithm of \citet{chen2023finite}, which combines principles of value iteration and best-response dynamics. Our primary contribution lies in introducing Tsallis entropy regularization for strategy updates, replacing the softmax smoothing (Shannon entropy) used in previous work.


\paragraph{Algorithm statement.} The pseudo-code of \algo is presented in Algorithm \ref{algorithm:TBRVI}. It takes as input the number of episodes $T$, the length of an episode $K$, and a regularization parameter $\eta$.
\begin{algorithm}\caption{Tsallis-smoothed Best-Response Dynamics with Value Iteration}\label{algorithm:TBRVI} 
	\begin{algorithmic}[1]
		\STATE \textbf{Input:} Integers $K$ and $T$, real number $\eta >0$, matrices $v_0^i=\bm{0}\in\mathbb{R}^{|\mathcal{S}|}$, $q_{t,0}^i=\bm{0}\in\mathbb{R}^{|\mathcal{S}||\mathcal{A}^i|}$ for all $t$, strategies $\pi_{t,0}^i(a^i|s)=1/|\mathcal{A}^i|$ for all $(s,a^i)$ and $t$.
		\FOR{$t=0,1,\cdots,T$}
		\FOR{$k=0,1,\cdots,K-1$}
		\STATE Update strategies: $\forall s \in \mathcal{S}, i \in \{1,2\}: \pi_{t,k+1}^i(s)=\pi_{t,k}^i(s)+\beta_k(\ts(q_{t,k}^i(s,.))-\pi_{t,k}^i(s))$ \label{line:policy_update}
		\STATE Sample actions:
		$a_k^i\sim \pi_{t,k+1}^i(\cdot|s_k)$ \label{line:action_sampling}
            \STATE Observe: $s_{k+1}\sim P(\cdot\mid s_k,a_k^i,a_k^{-i})$ \label{line:transition}
		\STATE Update values, for all $(s,a^i)$:  \label{line:td_learning}
                \begin{align*}    		    
                    q_{t,k+1}^i(s,a^i) \!=\! q_{t,k}^i(s,a^i) \! +\! \alpha_k \Big(\mathcal{R}^i(s_k,a_k^i,a_k^{-i}) \!+\! \gamma v_t^i (s_{k+1}) \!-\!q_{t,k}^i(s_k,a_k^i)\Big) \mathds{1}_{\{s=s_k,a^i = a_k^i\}} \nonumber
    		\end{align*}
		\ENDFOR
		\STATE $v_{t+1}^i(s)=\pi_{t,K}^i(s)^\top q_{t,K}^i(s,.)$ for all $s\in\mathcal{S}$ and set $s_0=s_K$ \label{line:VIteration}
		\ENDFOR
		\STATE \textbf{Output:} $\pi_{T,K}^i$
	\end{algorithmic}
\end{algorithm} 

In Algorithm \ref{algorithm:TBRVI}, the value functions and the $q$-functions are initialized at zero. In line \ref{line:policy_update}, the strategies are updated according to a smoothed best response. The smoothing is based on Tsallis entropy regularization with $\alpha = 1/2$, see \citet{tsallis1988possible},
leading to the following strategy update: 
\begin{equation}\label{eq:tsallis_smoothing}
    \ts(q_{t,k}^i(s)) =\underset{w \in \Delta^{|\cA^i|}}{\arg \max }\left\langle w,q_{t,k}^i(s,.)\right\rangle+\frac{4}{\eta} \sum_j \sqrt{w_j},
\end{equation}
where $\eta >0$ is a learning rate. Note that without adding regularization, the update becomes a maximization of $\left\langle w,q_{t,k}^i(s,.)\right\rangle$ which is equal to $\mathbb{E}_{a^i \sim w} [q_{t,k}^i(s,.)]$, similar to the best-response dynamics \citep{hofbauer2006best}. Finally, \citet{zimmert2021tsallis} provided a closed form for the Tsallis update:
$\ts(q_{t,k}^i(s)) = (4 \eta^{-2} (q_{t,k}^i(s,a) - x )^{-2})_{a\in \cA}$, where the constant $x \in \bR$ is defined implicitly through the normalization constraint $\sum_{a\in \cA} 4\left(\eta\left(q_{t,k}^i(s,a)-x \right)\right)^{-2}=1$.

Next, in line \ref{line:action_sampling}, the action of player $i$ is sampled from strategy $\pi_{t,k+1}^i(.|s_k)$. Subsequently, in line \ref{line:transition}, the MDP transitions to state $s_{k+1}$ according to the transition dynamics. Line \ref{line:td_learning} depicts the $q$-functions' updates. Finally, in line \ref{line:VIteration}, the value functions are updated similarly to value-iteration. Note that, unlike episodic settings, the state is not reset at the end of an episode. 


The inner loop of \algo (lines \ref{line:policy_update} to \ref{line:td_learning}) must ensure the convergence of strategies and $q$-functions, and this presents a key challenge compared to past settings and analyses. The convergence of value updates in the outer loop, line \ref{line:td_learning}, typically requires strategies to visit all state actions sufficiently \citep{sutton2018reinforcement}. Past works ensure this requirement by assuming strong reachability and using Shannon entropy, through softmax smoothing. By introducing the Tsallis entropy, we provide better reachability bounds, as we will prove. Intuitively, Tsallis entropy entails more exploration of suboptimal actions (Lemma \ref{lem:margins}) and leads to faster mixing times (Lemma \ref{lem:mixing_time}).

\subsection{Theoretical statement}

Here, we show \algo's finite-time convergence to an approximate Nash equilibrium. Rather than strong reachability and uniform mixing-time assumptions, we consider a relaxed assumption. 

\begin{assumption}\label{as:diameter}
    There \emph{exists} a joint strategy $\pi_r=(\pi_r^i,\pi_r^{-i})$ such that $P_{\pi_r}$ induces a Markov chain which is irreducible and has a finite mixing-time.
\end{assumption}
The irreducibility here is strictly weaker than strong reachability (Definition \ref{as:strong_reachability}) as it only concerns a single strategy instead of any possible strategy. Similarly, the existence of a finite mixing time for the strategy $\pi_r$ relaxes the assumption of uniformly bounded mixing times (Definition  \ref{as:bounded_mixing}). For example, the two-player MDP discussed after Equation \eqref{eq:exp_stationary} satisfies Assumption \ref{as:diameter} with any strategy in which the second player chooses action $d$ with positive probability. However, this MDP does not satisfy strong reachability or uniformly bounded mixing times as we previously showed.

We are now ready to present our convergence result. This is based on several key characteristics of the reference strategy $\pi_r$ of Assumption \ref{as:diameter}. First, the reachability here is measured by the quantities $d_r=\min \left\{k \geq 0: P_{\pi_r}^k\left(s, s^{\prime}\right)>0, \forall\left(s, s^{\prime}\right)\right\}$ and $\mu_{r,\min} = \min_{s\in \cS} \mu_r (s)$, where $\mu_r$ is the stationary distribution of strategy $\pi_r$. The quantity $\mu_{r,\min}$ is guaranteed to be positive by Assumption \ref{as:diameter} and Equation \eqref{eq:Equivalent_Reachability}. Lastly, our convergence rate depends on the mixing time of $\pi_r$, characterized by the smallest $\rho_r\in (0,1)$ such that for all $k\geq 0, \; \max_{s\in\mathcal{S}}\left\|P_{\pi_r}^k(s,\cdot)-\mu_r(\cdot)\right\|_{\text{TV}}\leq 2\rho_r^k$.

\begin{theorem}[Nash Gap bound]\label{thm:NashGapBound}
    Assume that the players follow Algorithm \ref{algorithm:TBRVI} with the parameters $\alpha_k =  \alpha/(k+h)$ and $\beta_k= \beta/(k+h)$, where $\alpha,h>0$ and $\frac{\alpha}{h} < 1$. Choose $\frac{\beta}{\alpha} \leq \frac{c_\eta\ell_{\eta}^3 (1-\gamma)^2}{6272 \eta^3 |\mathcal{S}|A_{\max}^4}$, where $\ell_\eta = 1 / \left(\sqrt{A}+\frac{\eta}{2(1-\gamma)}\right)^2$ and $c_\eta=\mu_{r,\min} \ell_\eta^3$. Then, under Assumption \ref{as:diameter}, it holds for all $K\geq k_0$:
    \begin{equation*}
        \mathbb{E}[\textit{NG}(\pi_{T,K}^i,\pi_{T,K}^{-i})] \leq \frac{ c_1 |\mathcal{S}| A_{\max} T \eta }{ (1-\gamma)^3 }\left(\frac{ \gamma+1 }{2}\right)^{T-1} + \frac{ c_2|\mathcal{S}|^2A_{\max}^2 L_\eta \tau_K^2\alpha^{3/2} }{\alpha_{k_0} \beta(1-\gamma)^5}\frac{1}{\sqrt{K}} + \frac{c_3 \sqrt{A_{\max}}}{\eta (1-\gamma)^2},
    \end{equation*}
    where $k_0=\min \left\{k \geq 0 \mid k \geq \tau_k\right\}$, $\tau_K = t_{\ell_\eta,\beta_k}$,  $L_\eta = \frac{2\log(8|\mathcal{S}|)}{\ell_\eta^{2d_r}\mu_{r,\min} \log(1/\rho_r)} + 2$,  and $\{c_j\}_{1\leq j\leq 3}$ are numerical constants.
\end{theorem}
Observe that $\ell_\eta = \mathcal{O}(\eta^{-2})$, $\tau_K=\mathcal{O}(\log(K) \eta^{4 d_r})$, $L_\eta = \mathcal{O}(\eta^{4 d_r})$, and $c_\eta = \mathcal{O}(\eta^{-6})$, and thus, they have polynomial dependence  on $\eta$. This enables us to establish convergence to an $\epsilon$-approximate Nash equilibrium with polynomial sample complexity in $1/\epsilon$ as will be shown below.\\
Theorem \ref{thm:NashGapBound} provides an upper bound on the Nash gap along the iterates of our algorithm. The first term in the inequality represents the bias induced by the minimax value iteration (line \ref{line:VIteration} of Algorithm \ref{algorithm:TBRVI}) and the contraction property of the Bellman operator due to the discount factor. The second term captures the combined convergence error and variance of the inner loop. Its scaling as $1/\sqrt{K}$ is consistent with typical bounds from online learning. The third term corresponds to a regularization bias arising from the use of Tsallis entropy smoothing. By optimizing these three terms with respect to $\eta$, we establish the convergence of the iterates to an $\epsilon$-approximate Nash equilibrium. 

\begin{corollary}[Sample Complexity]\label{cor:SampleComplexity}
    Under Assumption \ref{as:diameter} and with $\eta = K^{1/(24 d_r + 32)}$, for any $\epsilon> 0$, \algo  achieves $\mathbb{E}[\textit{NG}(\pi_{T,K}^i,\pi_{T,K}^{-i})]\leq \epsilon\:\:$ for $K = \widetilde{\mathcal{O}}\left(1/\epsilon^{24 d_r + 32}\right)$ and $T = \widetilde{\mathcal{O}}(\log(1/\epsilon))$. Thus, \algo learns an $\epsilon$-approximate Nash equilibrium with a polynomial sample complexity in $1/\epsilon$. 
\end{corollary}
To our knowledge, this is the first result that establishes a polynomial sample complexity in $1/\epsilon$ without requiring strong reachability or uniform mixing time bounds.  

We further show that for a player using \algo, her iterates converge in finite time to the best response of an opponent playing a stationary strategy. 
\begin{corollary}[Rationality]\label{cor:rationality}
	Let player $-i$ follow a strategy $\pi^{-i}$. Assume there exists a strategy $\pi_i$ such that the joint strategy $(\pi^i, \pi^{-i})$ induces an irreducible Markov chain with a finite mixing time. Then for any $\epsilon>0$, if the player $i$ follows \algo with $\eta = K^{1/(24 d_r +32)}$, then $\max_{\hat{\pi}^i} v^i_{(\hat{\pi}^i,\pi^{-i})}(s_0) - v^i_{(\pi_{T,K}^i,\pi^{-i})}(s_0)\leq \epsilon \:\:$ for $K = \widetilde{\mathcal{O}}\left(1/\epsilon^{24 d_r + 32}\right)$ and $T = \widetilde{\mathcal{O}}(\log(1/\epsilon))$. 
\end{corollary} 
Observe that the assumption in the statement of Corollary \ref{cor:rationality} is equivalent to Assumption \ref{as:diameter} in the case of a fixed opponent. The above corollary establishes the so-called \emph{rationality} of the algorithm as referred in \cite{cen2021fast, chen2021sample,wei2021last}. 



\paragraph{Comparison with state-of-the-art.} Let us compare with the closest works addressing payoff-based decentralized learning of equilibria in zero-sum Markov games. 
In \citet{wei2021last}, a sample complexity of $\tilde{\mathcal{O}}\left(1/\epsilon^{8}\right)$ was derived for the weaker notion of average iterate convergence. Furthermore, the above work required strong reachability and proposed an algorithm that updated strategies and $q$ functions on separate timescales, necessitating a higher level of coordination between players. Specifically, their method involved players iteratively coordinating to fix their strategies while collecting batches of samples to estimate the value functions. 
For last-iterate convergence, \citet{chen2021sample} derived a $\widetilde{\bigO} (1/\epsilon^{5.5})$ sample complexity assuming strong reachability and uniformly bounded mixing times. Furthermore, their algorithm required communicating the entropy of the player's strategy to the opponent. Recently, \citet{cai2024uncoupled} proved a last-iterate guarantee with a rate of $\tilde{ \mathcal{O} }\left( L^{1/\xi}/ \epsilon^{9+\xi} \right)$ for any $\xi>0$ under the assumption of strong reachability. Relaxing this assumption, \citet{cai2024uncoupled} showed a so-called sample path convergence, implying that ``for all states that players visit often enough, players learn an approximate Nash strategy''. As proven in \citet{cai2024uncoupled}, this convergence notion is weaker than its average-iterate and last-iterate counterparts. Finally, past efforts on relaxing strong reachability include \citep{sayin2021decentralized,chen2023finite}. Specifically, \citet{sayin2021decentralized} proved asymptotic convergence to a bias Nash equilibrium under a condition similar to Assumption \ref{as:diameter}, and \citet{chen2023finite} extended the previous result, under Assumption \ref{as:diameter}, to a finite-time convergence to a biased Nash equilibrium.


\section{Proof sketch}

Our objective in this section is to provide insight into the proof, highlighting the role of Tsallis entropy in overcoming assumptions of strong reachability and uniformly bounded mixing times.
We attribute the prevalence of restrictive assumptions to the popular use of softmax smoothing in the literature. Specifically, softmax smoothing induces strategies that converge too rapidly. Thus, the probability of selecting suboptimal actions decays exponentially fast, hindering the convergence of value functions. Moreover, this decay influences the mixing time bounds, which grow inversely proportional to the level of exploration of the strategies \citep[Lemma 4]{chen2023finite}.

\subsection{Key properties of Tsallis entropy} 
\label{sec:key_properties_tsallis}


Tsallis entropy,  $\cH_\alpha (\pi) = \frac{1}{1-\alpha}(1-\sum_i \pi_i^\alpha)$,  $\alpha \in [0,1]$,  generalizes the Shannon entropy and the log-barrier potential as special cases for $\alpha \to 1$ and $\alpha \to 0$, respectively \citep{agarwal2017corralling, abernethy2015fighting}. The strategies induced by Tsallis entropy have a closed-form expression, see Equation \eqref{eq:tsallis_smoothing}. This expression allows us to establish the following crucial lower bound on the strategies deployed by Algorithm \ref{algorithm:TBRVI}. 


\begin{lemma}[Policy lower bound]
\label{lem:margins}
	For all $t,k\geq 0$, $s \in S, a^i \in \cA$, $i\in \{1,2\}$ the following lower bound on the strategies $\pi_{t,k}^i$ of \algo holds:
            $ \pi_{t,k}^i(a^i|s) \geq \ell_{\eta}$,
        where $\ell_\eta = 1 / \left(\sqrt{A}+\frac{\eta}{2(1-\gamma)}\right)^2$.
\end{lemma}
The proof of this lemma is provided in Appendix \ref{ap:misc}. The above ensures that strategies are lower bounded by a function of the regularization coefficient $\eta$. This bound is utilized in the proof of Theorem \ref{thm:NashGapBound} in two ways. First, we use it to prove a lower bound on the stationary distribution of the strategies of \algo, see Lemma \ref{lem:stationary_distribution_lower_bound} below; Second, we can bound the mixing time of the strategies. 

\begin{lemma}[Stationary distribution lower bound]
\label{lem:stationary_distribution_lower_bound}
 Define the strategy class $\Pi_\delta =\{\pi=(\pi^i,\pi^{-i})\mid \min_{s,a^i}\pi^i(a^i|s)>\delta_i, \: \text{ for } \: i=1,2\}$, where $\delta_i,\delta_{-i} \in (0,1)$. For any strategy $\pi \in \Pi_\delta$ with stationary distribution $\mu_\pi$, the following lower bound on $\mu_\pi$ holds: $ \mu_\pi(s) \ge \mu_{r,\min} \delta_1 \delta_{2}$.
\end{lemma}

The proof of this lemma is provided in Appendix \ref{ap:misc}. Intuitively, it follows from the observation that a sufficiently exploratory strategy coincides with $\pi_r$ with some probability. This exploratory behavior is guaranteed by the use of Tsallis entropy regularization. The argument then involves algebraic manipulations and relating the stationary distribution of any exploratory strategy to $\mu_r$. Notably, Lemma \ref{lem:stationary_distribution_lower_bound} also holds for Shannon entropy-regularized strategies; however, the resulting lower bound would be exponentially small, making it impractical for the final analysis. 

Using the above-established properties, we now proceed to bound the Nash gap through drift inequalities defined below.

\subsection{Drift inequalities} 

The Nash gap can be decomposed as below \citep[Equation 7]{chen2023finite}.
\begin{align}\label{eq:nash_gap_decomposition}
	\nag(\pi_{T,K}^i,\pi_{T,K}^{-i}) &\leq C_0 \bigg( \mathcal{D}_\pi(T,K) + 2 \underbrace{\|v_T^i+v_T^{-i}\|_\infty}_{\mathcal{D}} +\sum_{i=1,2} \underbrace{\|v^i_T-v^i_{*}\|_\infty}_{\mathcal{D}_{i}} + \underbrace{\frac{8 \sqrt{A_{\max}}}{\eta}}_{\text{bias}} \bigg),
\end{align}
where $C_0$ is a constant, $\mathcal{D}_\pi(\cdot)$ is a term we refer to as strategy drift below,   and $v^i_{*}$ is the unique fixed-point of a Bellman operator (see Appendix \ref{app:notations}).
Our proof builds on that of \citet{chen2023finite}, establishing drift inequalities for each term on the right-hand side. Drift inequalities are inequalities showing a negative drift of the iterate (similar to Lyapunov inequalities) with additional terms arising from couplings with other iterates. Let us now highlight how we use Tsallis entropy properties established in Section \ref{sec:key_properties_tsallis} to establish drift inequalities for $\mathcal{D}_\pi$, $\mathcal{D}$ and $\mathcal{D}_i$, $i=1,2$. 


\paragraph{Policy drift $\cD_\pi$.} The first term  $\mathcal{D}_\pi$, in the Nash gap decomposition is the sum of the so-called Lyapunov functions \citep{hofbauer2005learning}, as defined below. 
\begin{align}\label{def:lyapunov_policy}
    \mathcal{D}_{\pi} = & \sum_s V_{v,s}\left(\pi^i, \pi^{-i}\right),  \nonumber \\
    V_{v,s}\left(\pi^i, \pi^{-i}\right) = & \!\!\sum_{i=1,2} \max_{\hat{\pi}^i \in \Delta^{\left|\mathcal{A}^i\right|}} \!\Big\{ \!\left(\hat{\pi}^i \!-\! \pi^i\right)^{\top} \!\mathcal{T}^i(v^i)(s) \pi^{-i} + \frac{1}{\eta} \left( \cH \left(\hat{\pi}^i\right) - \cH \left(\pi^i\right)\right)\!\Big\},
\end{align}
where $\mathcal{T}^i(v)(s,a^i,a^{-i}) \eqdef \mathcal{R}^i(a,a^i,a^{-i})+\gamma\mathbb{E}\left[ v(s_1)\mid s,a^i,a^{-i}\right]$. The function $V_{v,s}$ serves as a regularized Nash gap for the matrix game with payoffs $\mathcal{T}^i(v)(s,\cdot,\cdot)$. It is an adaptation of the Lyapunov function provided in \citet{hofbauer2005learning} for best response dynamics in matrix games. In particular, it is adjusted to accommodate the Markov games setting by using $\mathcal{T}^i(v)$ as the payoff matrices. Furthermore, rather than the previously adopted Shannon entropy, the regularization $\cH(\cdot)$ is with Tsallis entropy, aligning with our algorithmic choices. 

To analyze $\cD_{\pi}$, we prove in Lemma \ref{lem:properties_Lyapunov} that $V_{v,s}$ is strongly convex and smooth. A key technical contribution that we needed to prove these properties is the following Lipschitzness Lemma.

\begin{lemma}[Lipschitzness of Tsallis entropy]
\label{lem:lipschitz_tsallis}
    For all $\bm{R}$ and $\bm{R}' \in \bR^n$, we have:
    \begin{equation*}
         \quad \|\ts(\bm{R})- \ts(\bm{R}')\|_{2} \le 2 \sqrt{2} \eta \: n  \|\bm{R}- \bm{R}'\|_2.
    \end{equation*}
\end{lemma}
The proof of this lemma is provided in Appendix \ref{ap:Tsallis}. As a consequence, we show in Lemma \ref{lem:properties_Lyapunov} that $V_{v,s}$ is smooth and strongly convex, properties that are crucial for establishing the negative drift of $\cD_\pi$. Notably, an analogous Lipschitz property for Shannon entropy was derived in \citet{gao2017properties} and has been instrumental in prior convergence analyses, see \citet[Lemma A.7]{chen2023finite} and \citet[Lemma 6]{chen2021sample}. Given this significance, we believe the Lipschitz property of Tsallis entropy may also be of independent interest.


\paragraph{Value function drifts $\cD, \cD_i$, $i=1,2$.} We refer to $\cD$ and $(\mathcal{D}_{i})_{i=1,2}$ as value function drifts, as they relate to different terms in the estimation of value function $((v_t^i)_{i=1,2})_{t\ge 0}$. For each of these terms, we prove an inequality that exhibits a negative drift with an addition of a coupling term. The negative drift primarily arises from the contractiveness of the Bellman operator, while the coupling terms reflect the dependence on the value function of player $i$ on that of the other player, and the strategy iterates. The analysis of $\mathcal{D}_1$, $\mathcal{D}_2$, and $\mathcal{D}$ relies on properties of Tsallis entropy. 
For the convergence of value functions in line \ref{line:td_learning} of Algorithm \ref{algorithm:TBRVI} it is crucial that the strategies appear stationary. To ensure this, we leverage the derived upper bound on the mixing time of induced Markov chains, to determine a suitable episode length. A sufficiently long episode guarantees that the Markov chain is close to its stationary distribution, thus ensuring accurate value function estimation. 


\paragraph{Bias term.} The bias term  $\left(8 \sqrt{A_{\max}} / \eta\right)$, is the last term in our decomposition. The objective is to select a sufficiently large $\eta$ to control this regularization bias. Observe that this bias is also the last term in the Nash gap bound of Theorem \ref{thm:NashGapBound}, whereas the first two terms therein capture the cumulative effects of the drift terms $\cD_\pi, \cD, \cD_1, \cD_2$. Importantly, our bound on the drift functions scales polynomially with $\eta$ (see Theorem \ref{thm:NashGapBound}), allowing us to optimize the Nash gap bound with respect to $\eta$, effectively removing the bias and establishing a polynomial sample complexity in $1/\epsilon$. 

In contrast, prior approaches using softmax smoothing suffer from exponentially decaying strategy lower bounds and exponentially increasing mixing-time upper bounds, see \citet{auletta2013mixing}. As a result, softmax-smoothed algorithms cannot balance the bias term with the drift terms while maintaining a polynomial convergence rate in $1/\epsilon$. This underscores the critical advantage of Tsallis entropy, which yields a bias term that can be successfully controlled.

\subsection{Discussion}

\textbf{Tightness of bounds.} The sample complexity in our setting is of order $\tilde{\mathcal{O}}\left(1/\epsilon^{24 d_r + 32}\right)$, exhibiting an exponential dependence on $d_r$. We argue that this dependence is implicitly present in prior results and may be an inherent complexity of the problem. Specifically, strong reachability is characterized by the $L$ constant (see Definition \ref{as:strong_reachability}) in place of $d_r$. Existing convergence rates depend polynomially on $L$, however, we argue that $L$ itself can be exponentially large in $d_r$. Specifically, it is well established that for any $s,s' \in \cS$, the expected time to reach state $s'$ from state $s$ under strategy $\pi$, denoted as $T_{s\to s'}^\pi$, can be lower bounded by the mixing time, see \citet[Theorem 10.22]{durrett2019probability}, \citet[Theorem 1.3]{oliveira2012mixing}, and \citet[Theorem 5.7]{lovasz1995mixing}. Since $L$ serves as an upper bound on $T_{s\to s'}^\pi$ across all states and strategies, then $L$ is also lower bounded by the mixing time. Moreover, as established in Lemma \ref{lem:mixing_time}, the mixing time itself grows exponentially with $d_r$. We conclude that $L$ grows exponentially with $d_r$ and so do the convergence rates in past work. \\
In a single agent setting, \citet{auer2008near} established tight bounds showing that the dependence on quantities analogous to $L$ (diameter) is unavoidable. This result strongly suggests that the exponential dependence on $d_r$ is inherent to the problem. However, the constants in Corollary \ref{cor:SampleComplexity} perhaps could still be improved with refined analyses or alternative approaches to the problem.



\textbf{Implementability of the algorithm.} In Theorem \ref{thm:NashGapBound} and Corollary \ref{cor:SampleComplexity}, key quantities like $d_r$, $\rho_r$, and $\mu_{r,\min}$ are required to appropriately tune the hyperparameters of \algo. However, these quantities are typically unknown in practice and must be estimated. In single-agent reinforcement learning, several well-established methods exist for approximating such parameters. For instance, the mixing time constant $\rho_r$ can be estimated using coupling methods, as demonstrated in \citet[Section 4.2]{jerrum2003counting}. Similarly, practical techniques for estimating the stationary distribution of a Markov chain have been developed by \citet[Chapter IV]{asmussen2007stochastic}. On the other hand, while the estimation of $d_r$ has not yet been addressed in the literature, we believe it can be approximated using a simple heuristic: running a uniform strategy and tracking the minimum time required to transition between any pair of states. This approach, though elementary, could serve as a practical method for obtaining empirical estimates of $d_r$.



\textbf{Learning rate.} The learning rate $\beta_k$ satisfies $\beta_k = \beta /(k+h) \le  \frac{ \alpha c_\eta\ell_{\eta}^3 (1-\gamma)^2}{6272 \eta^3 |\mathcal{S}|A_{\max}^4 (k+h)}$, which means that $\beta_k = \mathcal{O}(\eta^{-12}/k)$, which can be very small in practice. While this is a necessary aspect of the theoretical analysis, it can be limiting in practice. Small learning rates of this nature are common in analyses of algorithms involving simultaneous value and strategy updates, both in single-agent reinforcement learning \citep{olshevsky2023small,konda1999actor,khodadadian2022finite} and in Markov games \citep{chen2023finite,cai2011minmax}. The approaches that collect many samples of the MDP per strategy update to estimate the values can avoid this issue \citep{wei2021last,cai2024uncoupled}. However, implementing such algorithms in a game setting requires additional coordination, as the value function of player $i$ depends on the strategy $\pi^{-i}$ of the opponent. Hence, the players need to fix their strategies in a coordinated way.

\section{Conclusion}

This work addressed learning an approximate Nash equilibrium in zero-sum Markov games by proposing a payoff-based and decentralized algorithm. The main contribution was relaxing strong reachability and uniform mixing time assumptions made in the prior works. Specifically, we introduced a relaxed reachability requirement in Assumption \ref{as:diameter} and proposed the \algo algorithm, which provably converges to an $\epsilon$-approximate Nash in polynomial time in $1/\epsilon$. The key algorithmic contribution was the use of Tsallis entropy to obtain smooth strategy updates.

 The work opens up several promising research directions. First, we plan to improve our sample complexity bounds using concentration inequality-based analyses, which have provided the best convergence rates so far under the assumptions of strong reachability and uniformly bounded mixing times \citep{wei2021last,chen2021sample}. 
Second, given the slow learning rate required in our algorithm, an important direction is to explore sample complexity lower bounds for Markov games under the payoff-based information and in the absence of coordination. Finally, to address real-world applications, extending both the approach and analysis to continuous state and action spaces, as well as validating them through real-world experiments, is a crucial next step.


  

\bibliographystyle{plainnat} 

\bibliography{main}

\appendix


\include{supplement}

\end{document}

%% file: supplement.tex
\section{Notations}\label{app:notations}
We dedicate this section to index all the notations used in this paper. Note that every notation is defined when it is introduced as well.

\renewcommand{\arraystretch}{1.5}
\begin{longtable}{l l l}
\caption{Notations}\label{tab:Notation}\\
\hline
 $T$ &$\eqdef$ & $K \times H$, total number of steps\\
 \midrule
 $ A_{\max}$ &$\eqdef$ & $\max \{|\cA^1|,|A^2|\}$.\\
 \midrule
 $v^i$ &$\eqdef$ & value function of agent $i \in \{1,2\}$, $\in\mathbb{R}^{|\mathcal{S}|}$\\
 \midrule
 $\mathcal{R}^i(a,a^i,a^{-i})$ &$\eqdef$ & Reward matrix for player $i (i \in \{1,2\})$\\
 \midrule
 $\mathcal{T}^i(v)(s,a^i,a^{-i})$ &$\eqdef$ & $\mathcal{R}^i(a,a^i,a^{-i})+\gamma\mathbb{E}\left[ v(s_1)\mid s_0=s,a_0^i=a^i,a_0^{-i}=a^{-i}\right]$\\
 \midrule
 $\textit{val}^i(X)$ &$\eqdef$ & $\max_{\pi^i\in\Delta^{|\mathcal{A}^i|}}\min_{\pi^{-i}\in\Delta^{|\mathcal{A}^{-i}|}}\{(\pi^i)^\top X\pi^{-i}\}$ \\
 &$=$ & $\min_{\pi^{-i}\in\Delta^{|\mathcal{A}^{-i}|}}\max_{\pi^i\in\Delta^{|\mathcal{A}^i|}}\{(\pi^i)^\top X\pi^{-i}\}$\\
 \midrule
 $\mathcal{B}^i(v)(s)$ &$\eqdef$ & $\textit{val}^i(\mathcal{T}^i(v)(s))$, the minimax Bellman operator \\
 \midrule
 $v_*^i$ &$\eqdef$ & The unique fixed-point of $\mathcal{B}^i(v^i)(s)$. Note that $v_*^i+v_*^{-i}=0$.\\
 \midrule
 $\cH(w)$ &$\eqdef$ & $ 4 \sum_{i \in [n]} \sqrt{w_i}$ for all $w \in \bR^n$, Tsallis-entropy \\
 \midrule
 $V_X\left(\pi^i, \pi^{-i}\right)$ &$\eqdef$ & $\sum_{i=1,2} \max _{\hat{\pi}^i \in \Delta^{\left|\mathcal{A}^i\right|}}\left\{\left(\hat{\pi}^i-\pi^i\right)^{\top}\! X_i \pi^{-i}+\frac{1}{\eta} \cH\left(\hat{\pi}^i\right)\! -\! \frac{1}{\eta} \cH\left(\pi^i\right)\right\}$ \\
 \midrule
 $\mathcal{D}_q(t,k)$ &$\eqdef$ & $\sum_{i=1,2}\|q_{t,k}^i-\bar{q}_{t,k}^i\|_2^2$ \\
 \midrule
 $v_{*, \pi^{-i}}^i(s)$ &$\eqdef$ & $\max _{\hat{\pi}^i} v_{\hat{\pi}^i, \pi^{-i}}^i(s)$\\
 \midrule
 $v_{\pi^i, *}^i$ &$\eqdef$ & $\min _{\hat{\pi}^{-i}} v_{\pi^i, \hat{\pi}^{-i}}^i(s)$\\
 \midrule
 $v_{\pi^{-i}, *}^{-i}(s)$ &$\eqdef$ & $\min _{\hat{\pi}^i} v_{\pi^{-i}, \hat{\pi}^i}^{-i}(s)$\\
 \midrule
 $v_{*, \pi^i}^{-i}(s)$ &$\eqdef$ & $\max _{\hat{\pi}^{-i}} v_{\hat{\pi}^{-i}, \hat{\pi}^i}^{-i}(s)$\\
 \midrule
 $\Pi_\delta$ &$\eqdef$ & $\{(\pi^i,\pi^{-i})\mid \forall i, \min_{s,a^i}\pi^i(a^i|s)>\delta_i \}$, where $\delta_i \in (0,1)$.\\
 \bottomrule
\end{longtable}

\section{Supporting lemmas}

In this section, we present technical results that are crucial for our proof. Particularly, we provide new results for the Tsallis entropy smoothing.

\subsection{Tsallis entropy}
\label{ap:Tsallis}

Recall Tsallis-entropy in an $n$-dimensional space:  $\cH_\alpha (\pi) = \frac{1}{1-\alpha}(1-\sum_i \pi_i^\alpha)$. In this paper, we consider Tsallis entropy with $\alpha = 1/2$, which can be equivalently written \citep{zimmert2021tsallis}: $\cH(w) \eqdef 4 \sum_{i \in [n]} \sqrt{w_i}$. In algorithm \ref{algorithm:TBRVI}, we use this entropy as a regularization to the strategy update to define the Tsallis smoothing:
\begin{equation*}
    \ts(\bm{R}) = \argmax_{w \in \Delta^n} \left\langle w, \bm{R}\right\rangle + \frac{1}{\eta} \cH(w), \qquad \text{for all } \bm{R} \in \bR^n 
\end{equation*}
where $\eta$ is a positive scalar. A closed form was provided in \citet{zimmert2021tsallis} for the Tsallis smoothing:
\begin{equation*}
    \ts(\bm{R})_i = 4 / \left(\eta\left(\bm{R}_i - x \right)\right)^2,
\end{equation*}
where $x \in \bR$ is defined implicitly through the normalization constraint $\sum_i 4\left(\eta\left(\bm{R}_{i}-x \right)\right)^{-2}=1$.

\begin{lemma}[Normalization factor]
\label{lem:normalization_factor_bound}
Let $\bm{R} \in \bR^n$, and let $x \in \bR$ be the normalization factor of $\ts(\bm{R})$, then we can show that
\begin{equation*}
    2/\eta  \le  x - \max_i \bm{R}_i \le 2\sqrt{n}/\eta
\end{equation*}
\end{lemma}
\begin{proof}
Let $i^\star = \argmax \bm{R}_{i}$, then we have $w_{i^\star}\in [1/n,1]$ (see remark below) and 
\begin{equation*}
   x = \bm{R}_{i^\star}- 2\eta^{-1} \ts(\bm{R})_{i^\star}^{-\frac{1}{2}} 
\end{equation*}
Since the function $y : \rightarrow 4(\eta(\bm{R}_i - y))^{-2}$ is monotonous in the domain $I = [\bm{R}_{i^\star} + 2/\eta, \bm{R}_{i^\star} + 2\sqrt{n}/\eta ]$,
there exists a unique value in $I$, such that $\sum_i 4\left(\eta_t \left(\bm{R}_i - x \right)\right)^{-2}=1$.
\end{proof}
\begin{remark}
    Note that there are $n+1$ candidates for the normalization factor, such that the smallest one is smaller than $\min_i \br_i$ and the second smallest one is smaller than the second smallest $(\br)_j$, and so on until the $n^{\text{th}}$ one, the $(n+1)^{\text{th}}$ candidate is larger than $\max_i \br_i$. The $(n+1)^{\text{th}}$ candidate is the normalization factor because it maximizes the objective in equation \eqref{eq:tsallis_smoothing}.
\end{remark}

\begin{lemma*}[Lipschitzness, Lemma \ref{lem:lipschitz_tsallis} in the main text]
    For all $\bm{R}$ and $\bm{R}' \in \bR^n$, we have:
    \begin{equation*}
         \quad \|\ts(\bm{R})- \ts(\bm{R}')\|_{2} \le 2 \sqrt{2} \eta \: n  \|\bm{R}- \bm{R}'\|_2
    \end{equation*}
    This means that Tsallis-smoothing is $2 \sqrt{2} \eta$-Lipschitz with respect to $\|.\|_2$.
\end{lemma*}
\begin{proof}
    Let $\bm{R}$ and $\bm{R}'$ be two vectors in $\bR^n$, and consider $\br^0, \ldots, \br^n \in \bR^n$ such that $\br^0 = \br$, and for $j\in \{1,\ldots, n\}$, we define $ \br^j \eqdef (\br'_1, \ldots, \br'_j, \br_{j+1},\ldots,\br_n)$, we also denote $x_0, \ldots, x_n$ their respective Tsallis-normalization factor.

    Let $j \in \{1,\ldots, n\}$ we have:
    
    \textbf{1) If $\br_j^{\prime} \ge \br_j$:} then $x_j > x_{j-1}$, this is because the function $y : \rightarrow \sum_i 4(\eta(\bm{R}_i - y))^{-2}$ is decreasing in the domain $[x_{j-1}, \max_i \bm{R}_i^j + 2\sqrt{K}/\eta ]$ and it is larger than $1$ in $x_{j-1}$.
    
    The latter entails that for all $l \neq j, \ts(\br^{j})_l \le \ts(\br^{j-1})_l$, which implies that $\ts(\br^{j-1})_j \ge \ts(\br^{j-1})_j$ to keep the normalization condition. Then
    \begin{align*}
        \ts(\br^{j-1})_j \ge \ts(\br^{j-1})_j &\implies \frac{1}{\left(x_j-\br_j^{\prime}\right)^2} \ge \frac{1}{\left(x_{j-1}-\br_j\right)^2},\\
        &\implies  x_j - x_{j-1} \le \br'_j - \br_j,
    \end{align*}
    the last line follows because $x_j \ge \br'_j$ and  $x_{j-1} \ge \br_j$ thanks to Lemma.~\ref{lem:normalization_factor_bound}.
    
    \textbf{2) If $\br_j^{\prime} \le \br_j$:} Similarly, we can prove that this entails that $x_j - x_{j-1} \ge \br'_j - \br_j$.
    
    \textbf{In 1) and 2)} we have shown that for all $j \in \{1,\ldots, n\}$ we have: $x_j - x_{j-1} \le \br'_j - \br_j$. Therefore, we can show
    \begin{equation*}
        \sum_{l=1}^n (\br'_l - \br_l) \mathds{1}_{\br'_l \le \br_l} \le x_n - x_0 \le \sum_{l=1}^n (\br'_l - \br_l) \mathds{1}_{\br'_l \ge \br_l}
    \end{equation*}
    
    Now that we have studied the variation of the normalization factor we are able to analyze the variation of the Tsallis-induced weights from a variation of the $\br$ vector. Let $j \in \{1,\ldots, n\}$, we have:
    \begin{align*}
        \ts(\bm{R})_j - \ts(\bm{R}')_j &= \frac{4}{(\eta (\bm{R}_j - x_0))^2} - \frac{4}{(\eta (\bm{R}'_j - x_n))^2} \\
        &= \frac{4}{\eta^2} \left(\frac{(\bm{R}'_j - x_n)^2 - (\bm{R}_j - x_0)^2}{(\bm{R}'_j - x_n)^2 (\bm{R}_j - x_0)^2}\right) \\
        &= \frac{4}{\eta^2} \left(\frac{(\bm{R}'_j - \bm{R}_j + x_0 - x_n)(\bm{R}_j - x_0 + \bm{R}'_j - x_n)}{(\bm{R}'_j - x_n)^2 (\bm{R}_j - x_0)^2}\right) \\
        &= \frac{4}{\eta^2} \frac{\bm{R}'_j - \bm{R}_j + x_0 - x_n}{(\bm{R}'_j - x_n) (\bm{R}_j - x_0)} \left( \frac{1}{\bm{R}'_j - x_n} + \frac{1}{\bm{R}_j - x_0} \right) \\
        &\le 8 \eta |\bm{R}'_j - \bm{R}_j + x_0 - x_n| \\
        &\le 8 \eta \|\bm{R}' - \bm{R}\|_1 
    \end{align*}
    where the penultimate inequality follows since for all $j$, $\frac{1}{\eta (\bm{R}'_j - x_n)} \le 1$ and $\frac{1}{\eta (\bm{R}_j - x_0)} \le 1$ because they are square-roots of probabilities.
    
    Finally: 
    \begin{equation*}
        \|\ts(\bm{R}) - \ts(\bm{R}') \|_2 \le 2 \eta  \sqrt{2n} \|\bm{R}' - \bm{R}\|_1 \le 2 \sqrt{2} \eta \: n \|\bm{R}' - \bm{R}\|_2
    \end{equation*}
    which concludes the proof.
\end{proof}

\subsection{Markov chain tools}

The following lemmas establish exploration properties under assumption \ref{as:diameter}. To present the results, we need additional notations. Under Assumption \ref{as:diameter}, there exists a joint strategy $\pi_r$ such that its induced Markov chain has a unique stationary distribution $\mu_r\in\Delta^{|\mathcal{S}|}$. Since the Markov chain induced by $\pi_r$ is irreducible, it satisfies $T_{s \mapsto s}^{\pi_r} < \infty$ for all $s\in S$. Therefore, Equation \ref{eq:Equivalent_Reachability} implies that the minimum component of $\mu_r$, which we denote as $\mu_{r,\min}$, is positive. In addition, it is shown in \citet{levin2017markov} that Assumption \ref{as:diameter} implies the existence of $\rho_r\in (0,1)$ such that $\max_{s\in\mathcal{S}}\left\|P_{\pi_r}^k(s,\cdot)-\mu_r(\cdot)\right\|_{\text{TV}}\leq 2\rho_r^k$ for all $k\geq 0$, where $P_{\pi_r}$ is the transition probability matrix of the Markov chain $\{S_k\}$ under $\pi_r$.

Finally, we restrict our attention to the strategy class $\Pi_\delta:=\{\pi=(\pi^i,\pi^{-i})\mid \min_{s,a^i}\pi^i(a^i|s)>\delta_i,\min_{s,a^{-i}}\pi^{-i}(a^{-i}|s)>\delta_{-i} \}$, where $\delta_i,\delta_{-i} \in (0,1)$. This strategy class is sufficient for our setting as it captures the policies deployed by our algorithm, see lemma \ref{lem:margins}.

\begin{lemma}[\cite{zhang2023global}, Lemma 4]\label{lem:zhang}
	Suppose that Assumption \ref{as:diameter} is satisfied. Then we have the following results.
	\begin{enumerate}
		\item For any $\pi=(\pi^i,\pi^{-i})\in \Pi_\delta$, the Markov chain $\{S_k\}$ induced by the joint strategy $\pi$ is irreducible and aperiodic, hence admits a unique stationary distribution $\mu_\pi\in\Delta^{|\mathcal{S}|}$.
		\item 
		Let $G:\mathbb{R}^{|\mathcal{S}|A_{\max}}\mapsto\mathbb{R}^{|\mathcal{S}|}$ be the mapping from a strategy $\pi\in \Pi_\delta$ to the unique stationary distribution $\mu_\pi$ of the Markov chain $\{S_k\}$ induced by $\pi$. Then $G(\cdot)$ is Lipschitz continuous with respect to $\|\cdot\|_\infty$, with Lipschitz constant $L_\delta:=\frac{2\log(8|\mathcal{S}|/\rho_\delta)}{\log(1/\rho_\delta)}$, where  $\rho_\delta=\rho_r^{(\delta_i\delta_{-i})^{d_r}\mu_{r,\min}}$ and $d_r:=\min\{k\geq 0\;:\; P_{\pi_r}^k(s,s')>0,\;\forall\;(s,s')\}$.
	\end{enumerate}
\end{lemma}


\begin{lemma}[\cite{chen2023finite}, Lemma 4.2]\label{lem:mixing_time}
    Suppose that Assumption \ref{as:diameter} is satisfied. Then, it holds that $\sup_{\pi\in \Pi_\delta}\max_{s\in\mathcal{S}}\|P_\pi^k(s,\cdot)-\pi_\pi(\cdot)\|_{\text{TV}}\leq 2\rho_\delta^k$ for any $k\geq 0$, where  $\rho_\delta=\rho_r^{(\delta_i\delta_{-i})^{d_r}\mu_{r,\min}}$ and $d_r:=\min\{k\geq 0\;:\; P_{\pi_r}^k(s,s')>0,\;\forall\;(s,s')\}$. As a result, we have
    \begin{align}\label{eq:mixing_time_definition}
        \sup_{\pi\in \Pi_\delta}t_{\pi,\epsilon}\leq \frac{t_{\pi_r,\epsilon}}{(\delta_i\delta_{-i})^{d_r}\mu_{r,\min}},
    \end{align}
    where we recall that $t_{\pi,\epsilon}$ is the $\epsilon$ -- mixing time of the Markov chain $\{S_k\}$ induced by $\pi$.
\end{lemma}

This enables us to see the explicit dependence of the mixing time on the policies' lower bounds $\delta_i$, $\delta_{-i}$, and the mixing time of the benchmark exploration strategy $\pi_r$. Note that, as the margins $\delta_i,\delta_{-i}$ approach zero, the uniform mixing time in Lemma \ref{lem:mixing_time} goes to infinity. It was demonstrated in \cite{chen2023finite} that the mixing time of almost deterministic policies does diverge to infinity in simple MDP examples.  

Given Lemma \ref{lem:mixing_time}, we have fast mixing for all policies in $\Pi_\delta$ if the margins $\delta_i,\delta_{-i}$ are large, the mixing time of $\pi_r$ is small, and the stationary distribution is balanced (large $\pi_{b,\min}$).

\textbf{Notation} For simplicity, when $\Pi_\delta=\Pi_{\ell_\eta}$, we denote $\rho_\eta:=\rho_\delta$, and $L_\eta:=L_\delta$. We also define $c_\eta:=\mu_{r,\min} \ell_\eta^3$.

\subsection{Miscellaneous}
\label{ap:misc}

Here we present the proof of the lower bound on the policies deployed in algorithm \ref{algorithm:TBRVI}, and recall a lemma that bounds the value functions.

\begin{lemma*}[Lemma \ref{lem:margins} in the main text]
	It holds for all players $i \in \{1,2\}$, times $t,k\geq 0$ and state-action pairs $(s,a^i,a^{-i})$ that
    \begin{equation*}
        \pi_{t,k}^i(a^i|s) \geq \ell_{\eta} \eqdef 1 / \left(\sqrt{A}+\frac{\eta}{2(1-\gamma)}\right)^2,
    \end{equation*}
    we call this the margin property of the deployed policies.
\end{lemma*}

\begin{proof}
    We recall the policy-update equation:
    \begin{equation*}
        \pi_{t+1}^i=\pi_t^i+\beta_k\left(\ts\left(q_k^i\right)-\pi_t^i\right).
    \end{equation*}
    By lemma \ref{lem:normalization_factor_bound} we have that $2/\eta  \le  x - \max_i \bm{R}_i \le 2\sqrt{n}/\eta$, and by Lemma\ref{lem:bounded_value_function} we know that $\|q_{t,k}^i\|_\infty\leq \frac{1}{1-\gamma}$ therefore:
    \begin{equation*}
        \forall s \in \cS, a \in \cA: \ts\left(q_k^i\right)(a) \ge  \frac{1}{(\sqrt{A_{\max}}+\frac{\eta}{2(1-\gamma)})^2}.
    \end{equation*}    
    Next, it is easy to show by induction that $\pi_{t,k}^i(a|s) \geq \ell_{\eta}$ for all $s \in \cS, a \in \cA$. Indeed, $\pi_{t,0}^i(a|s) = 1/A_{\max} \geq \ell_{\eta}$, and for a given $k \ge 0$ if  $\pi_{t,k}^i(a|s) \geq \ell_{\eta}$ then:
    \begin{align*}
        \pi_{t,k+1}^i(a|s) &= \pi_t^i(a|s) \ell_{\eta}+\beta_k\left(\ts\left(q_k^i\right)_a -\pi_t^i(a|s)\right) \\
        &= (1-\beta_k)\underbrace{\pi_t^i(a|s)}_{\geq \ell_{\eta}} + \beta_k \underbrace{\ts\left(q_k^i\right)_a }_{\geq \ell_{\eta}} \\
        &\ge (1-\beta_k)\ell_{\eta} + \beta_k \ell_{\eta} = \ell_{\eta}
    \end{align*}
    which concludes the proof.
\end{proof}

\begin{lemma*}[Lemma \ref{lem:stationary_distribution_lower_bound} in the main text]
 Consider $\delta_i,\delta_{-i} \in (0,1)$, and a strategy $\pi \in \Pi_\delta$ with stationary distribution $\mu_\pi$. Then, the following lower bound on $\mu_\pi$ holds: $\mu_\pi(s) \ge \mu_{r,\min} \delta_1 \delta_{2}$.
\end{lemma*}

\begin{proof}
    Fix $s \in \cS$, to lower bound $\mu_{t,k}^i(s)$, notice that for any $\pi\in \Pi_{\delta}, k\in \bN$, we have:
     \begin{align*}
         \mu_\pi(s') &= \sum_s \bP_\pi^k (s' |s) \mu_\pi(s) = \sum_s \left(\sum_a (P^k(s' |s,a) \pi(a)\right) \mu_\pi(s) \\
         &= \sum_s \left(\sum_a (P^k(s' |s,a) \frac{\pi (a)}{\pi_r(a)} \pi_r(a)\right) \mu_\pi(s) \\
         &\ge \delta_1 \delta_{2} \sum_s \left(\sum_a (P^k(s' |s,a) \pi_r(a)\right) \mu_\pi(s) \\
         &\ge \delta_1 \delta_{2} \sum_s  \bP_{\pi_r}^k (s' |s) \mu_\pi(s)
     \end{align*}
    where the third line follows because $\pi_r(s)\le 1 \le \frac{1}{\delta_1 \delta_{2}} \pi(s)$. Now the inequality above holds for any $k \in \bN$, and in the limit $k\to\infty$ we have that $\bP_{\pi_r}^k (s' |s) \to \mu_r(s')$, therefore:
    \begin{equation*}
        \mu_\pi(s') \ge \delta_1 \delta_{2} \sum_s  \mu_r (s') \mu_\pi(s) \ge \delta_1 \delta_{2} \mu_{\pi_r, \min},
    \end{equation*}
    which concludes the proof. 
\end{proof}

\begin{lemma}[Lemma A.1 of \cite{chen2023finite}]\label{lem:bounded_value_function}
	It holds for all $t,k\geq 0$ and $i\in \{1,2\}$ that 
	\begin{enumerate}
		\item $\|v_t^i\|_\infty\leq \frac{1}{1-\gamma}$,
		\item $\|q_{t,k}^i\|_\infty\leq \frac{1}{1-\gamma}$.
	\end{enumerate} 
\end{lemma}

\section{Sample complexity analysis}
\label{ap:sample_complexity_analysis}


Before presenting our proof, we highlight its structure closely follows from \cite{chen2023finite}. Our primary contribution lies in incorporating Tsallis-entropy and adapting the affected arguments within the existing framework. Accordingly, we provide full proofs only for novel results or those requiring adaptation, while results directly from \cite{chen2023finite} are cited without proofs.

We begin the analysis by restating the Nash gap decomposition (\cf Equation \ref{eq:nash_gap_decomposition}):
\begin{equation*}
	\nag(\pi_{T,K}^i,\pi_{T,K}^{-i})\!\leq\! C_0\bigg(2\|v_T^i+v_T^{-i}\|_\infty\!+\!\sum_{i=1,2}\|v^i_T-v^i_{*}\|_\infty\!+\!\mathcal{D}_\pi(T,K)\!+\!\frac{8}{\eta} \sqrt{A_{\max}}\bigg),
\end{equation*}
where $C_0$ is a constant, and $\mathcal{D}_\pi(\cdot)$ is function of the policies at times $T$ and $K$ (\cf Equation \ref{def:lyapunov_policy}). This decomposition was proved in \citet[Lemmas A.3, A.4]{chen2023finite}.

We define some requirements for the parameters $\alpha_k$ and $\beta_k$. For simplicity of notation, given $k_1\leq k_2$, we denote $\alpha_{k_1,k_2}=\sum_{k=k_1}^{k_2}\alpha_k$, and for $k\geq 0$, we define $\tau_k=t_{\ell_\eta,\beta_k}$ where $\ell_\eta$ is the lower bound on policies, defined in Lemma \ref{lem:margins}. Observe that when $\beta_k = \mathcal{O}(1/k)$, then $\tau_k=\mathcal{O}(\log(k))$ due to the geometric mixing property in Lemma \ref{lem:mixing_time}.

We require learning rates such that $\alpha_{k-\tau_k,k-1}\leq 1/4$ for all $k\geq \tau_k$ and $\beta / \alpha\leq \frac{c_\eta\ell_{\eta}^3 (1-\gamma)^2}{6272 \eta^3 |\mathcal{S}|A_{\max}^4}$. When using diminishing stepsizes $\alpha_k=\frac{\alpha}{k+h}$ and $\beta_k=\frac{\beta}{k+h}$, we additionally require $\beta>2$.



The above requirements are necessary for our proof of convergence in Theorem \ref{thm:NashGapBound}. They can be explicitly satisfied since $\tau_k=\mathcal{O}(\log(1/k))$ with the chosen parameters. Finally, the parameter $k_0$ that appears in Theorem \ref{thm:NashGapBound} is defined to be $\min\{k\geq 0\mid k\geq \tau_k \}$.

\subsection{Outer-loop}



\begin{lemma}[Nash Gap in terms of value iterates]\label{lem:Nash_Gap_Decomp}
	It holds for all $t\geq 0$ and $i=1,2$ that
	\begin{align*}
		\left\|v^i_{*,\pi_{t,K}^{-i}}-v^i_{\pi_{t,K}^i,\pi_{t,K}^{-i}}\right\|_\infty \leq \frac{2}{1-\gamma}\bigg(&2\|v_t^i+v_t^{-i}\|_\infty+2\|v^i_t-v^i_{*}\|_\infty + \max_{s}V_{v_t,s}(\pi_{t,K}^i(s),\pi_{t,K}^{-i}(s)) \\
        & +\frac{8}{\eta} \sqrt{A_{\max}}\bigg).
	\end{align*}
\end{lemma}

\begin{proof}
    This proof is adapted from the proof of Lemma A.4 of \cite{chen2023finite}.
    
    For any $t\geq 0$, $s\in\mathcal{S}$, and $i\in \{1,2\}$, we have
    \begin{align*}
    	\left|v^i_{*,\pi_{t,K}^{-i}}(s)-v^i_{\pi_{t,K}^i,\pi_{t,K}^{-i}}(s)\right| &= v^i_{*,\pi_{t,K}^{-i}}(s)-v^i_{\pi_{t,K}^i,\pi_{t,K}^{-i}}(s) \\
    	&\leq v^i_{*,\pi_{t,K}^{-i}}(s)-v^i_{\pi_{t,K}^i,*}(s) \\
    	&= - v^{-i}_{\pi_{t,K}^{-i},*}(s)-v^i_{\pi_{t,K}^i,*}(s) \\
            &= v^i_*(s)-v^{-i}_{\pi_{t,K}^{-i},*}(s)+v^{-i}_*(s)-v^i_{\pi_{t,K}^i,*}(s) \\
    	&\leq \|v^{-i}_*-v^{-i}_{\pi_{t,K}^{-i},*}\|_\infty+\|v^i_*-v^i_{\pi_{t,K}^i,*}\|_\infty. \numberthis \label{eq:last_policy_bound}
    \end{align*}
    We now bound the two terms on the r.h.s above. For the first term, note that for any $s\in\mathcal{S}$ and $t\geq 0$, we have
    \begin{align*}
        0 \leq v^{-i}_*(s)-v^{-i}_{\pi_{t,K}^{-i},*}(s) = & v^i_{*,\pi_{t,K}^{-i}}(s)-v^i_*(s) \\
        = & \max_{\pi^i}(\pi^i)^\top  \mathcal{T}^i(v^i_{*,\pi_{t,K}^{-i}})(s)\pi_{t,K}^{-i}(s)-\max_{\pi^i}\min_{\pi^{-i}}(\pi^i)^\top \mathcal{T}^i(v_*^i)(s)\pi^{-i} \\
    	= & |\max_{\pi^i}(\pi^i)^\top  \mathcal{T}^i(v^i_{*,\pi_{t,K}^{-i}})(s)\pi_{t,K}^{-i}(s)-\max_{\pi^i}(\pi^i)^\top  \mathcal{T}^i(v^i_{*})(s)\pi_{t,K}^{-i}(s)| \\
    	& + |\max_{\pi^i}(\pi^i)^\top  \mathcal{T}^i(v^i_{*})(s)\pi_{t,K}^{-i}(s)-\max_{\pi^i}\min_{\pi^{-i}}(\pi^i)^\top \mathcal{T}^i(v_*^i)(s)\pi^{-i}| \\
    	\leq & \max_{\pi^i}|(\pi^i)^\top  (\mathcal{T}^i(v^i_{*,\pi_{t,K}^{-i}})(s)-\mathcal{T}^i(v^i_{*})(s))\pi_{t,K}^{-i}(s)| \\
    	& +|\max_{\pi^i}(\pi^i)^\top  \mathcal{T}^i(v^i_{*})(s)\pi_{t,K}^{-i}(s)-\max_{\pi^i}\min_{\pi^{-i}}(\pi^i)^\top  \mathcal{T}^i(v^i_t)(s)\pi^{-i}| \\
    	& +|\max_{\pi^i}\min_{\pi^{-i}}(\pi^i)^\top  \mathcal{T}^i(v^i_t)(s)\pi^{-i}-\max_{\pi^i}\min_{\pi^{-i}}(\pi^i)^\top \mathcal{T}^i(v_*^i)(s)\pi^{-i}| \\
    	\leq & \underbrace{\max_{\pi^i}|(\pi^i)^\top  (\mathcal{T}^i(v^i_{*,\pi_{t,K}^{-i}})(s)-\mathcal{T}^i(v^i_{*})(s))\pi_{t,K}^{-i}(s)|}_{\hat{E}_1} \\
    	& + \underbrace{|\max_{\pi^i}(\pi^i)^\top  \mathcal{T}^i(v^i_{*})(s)\pi_{t,K}^{-i}(s)-\max_{\pi^i}(\pi^i)^\top  \mathcal{T}^i(v^i_t)(s)\pi_{t,K}^{-i}(s)|}_{\hat{E}_2} \\
    	& + \underbrace{\max_{\pi^i}(\pi^i)^\top  \mathcal{T}^i(v^i_t)(s)\pi_{t,K}^{-i}(s)-\max_{\pi^i}\min_{\pi^{-i}}(\pi^i)^\top  \mathcal{T}^i(v^i_t)(s)\pi^{-i}}_{\hat{E}_3} \\
    	& + \underbrace{|\max_{\pi^i}\min_{\pi^{-i}}(\pi^i)^\top  \mathcal{T}^i(v^i_t)(s)\pi^{-i}-\max_{\pi^i}\min_{\pi^{-i}}(\pi^i)^\top \mathcal{T}^i(v_*^i)(s)\pi^{-i}|}_{\hat{E}_4}. \numberthis \label{eq:connect1}
    \end{align*}
    We next bound the terms $\{\hat{E}_j\}_{1\leq j\leq 4}$. For any $v_1^i,v_2^i\in\mathbb{R}^{|\mathcal{S}|}$, we have for any $(s,a^i,a^{-i})$ that
    \begin{align*}
        |\mathcal{T}^i(v_1^i)(s,a^i,a^{-i})-\mathcal{T}^i(v_2^i)(s,a^i,a^{-i})| &= \gamma|\mathbb{E}[v_1^i(s_1)-v_2^i(s_1)\mid s_0=s,a_0^i=a^i,a_0^{-i}=a^{-i}]|\\
        &\leq \gamma\|v_1^i-v_2^i\|_\infty,
    \end{align*}
    which implies that $\|\mathcal{T}^i(v_1^i)-\mathcal{T}^i(v_2^i)\|_\infty\leq \gamma\|v_1^i-v_2^i\|_\infty$.
    As a result, we have
    \begin{align*}
    	\hat{E}_1\leq\;& \|\mathcal{T}^i(v^i_{*,\pi_{t,K}^{-i}})-\mathcal{T}^i(v^i_{*})\|_\infty\leq \gamma\|v^i_{*,\pi_{t,K}^{-i}}-v^i_{*}\|_\infty,\\
    	\hat{E}_2\leq \;&\|\mathcal{T}^i(v_t^i)-\mathcal{T}^i(v^i_{*})\|_\infty\leq \gamma\|v^i_t-v^i_{*}\|_\infty,\\
    	\hat{E}_4\leq\;& \|\mathcal{T}^i(v_t^i)-\mathcal{T}^i(v^i_{*})\|_\infty\leq \gamma\|v^i_t-v^i_{*}\|_\infty.
    \end{align*}
    Finally, to bound $\hat{E}_3$, observe that
    \begin{align*}
    	\hat{E}_3 \leq &\left|\max_{\pi^i}(\pi^i)^\top \mathcal{T}^i(v_t^i)(s)\pi_{t,K}^{-i}(s)-\min_{\pi^{-i}}\pi_{t,K}^i(s)\mathcal{T}^i(v_t^i)(s)\pi^{-i}\right|\\
    	\leq &\left|\max_{\pi^{-i}}(\pi^{-i})^\top  \mathcal{T}^{-i}(v_t^{-i})(s)\pi_{t,K}^i(s)+\min_{\pi^{-i}}(\pi^{-i})^\top \mathcal{T}^i(v_t^i)(s)^\top \pi_{t,K}^i(s)\right| \\
        & \quad + \left|\sum_{i=1,2}\max_{\pi^i}(\pi^i)^\top \mathcal{T}^i(v_t^i)(s)\pi_{t,K}^{-i}(s)\right|\\
    	\leq &\left|\max_{\pi^{-i}}(\pi^{-i})^\top  \mathcal{T}^{-i}(v_t^{-i})(s)\pi_{t,K}^i(s)-\max_{\pi^{-i}}(\pi^{-i})^\top [-\mathcal{T}^i(v_t^i)(s)]^\top \pi_{t,K}^i(s)\right| \\
        & \quad + \left|\sum_{i=1,2}\max_{\pi^i}(\pi^i)^\top \mathcal{T}^i(v_t^i)(s)\pi_{t,K}^{-i}(s)\right|\\
    	\leq &\max_{\pi^{-i}}\left|(\pi^{-i})^\top  (\mathcal{T}^{-i}(v_t^{-i})(s)+[\mathcal{T}^i(v_t^i)(s)]^\top) \pi_{t,K}^i(s)\right| + \left|\sum_{i=1,2}\max_{\pi^i}(\pi^i)^\top \mathcal{T}^i(v_t^i)(s)\pi_{t,K}^{-i}(s)\right|\\
    	\leq &\max_{a^i,a^{-i}}\left|\mathcal{T}^i(v_t^i)(s,a^i,a^{-i})+\mathcal{T}^{-i}(v_t^{-i})(s,a^i,a^{-i})\right| +\left|\sum_{i=1,2}\max_{\pi^i}(\pi^i)^\top \mathcal{T}^i(v_t^i)(s)\pi_{t,K}^{-i}(s)\right|\\
    	\leq \;&\gamma\|v_t^i+v_t^{-i}\|_\infty+\left|\sum_{i=1,2}\max_{\pi^i}(\pi^i)^\top \mathcal{T}^i(v_t^i)(s)\pi_{t,K}^{-i}(s)\right|,
    \end{align*}
    where the last line follows because:
    \begin{align*}
        |\mathcal{T}^i(v_t^i)(s,a^i,a^{-i})+\mathcal{T}^{-i}(v_t^{-i})(s,a^i,a^{-i})| &= \gamma \bigg|\mathbb{E}[v_t^i(S_1)+v_t^i(S_1)\mid s_0=s,a_0^i=a^i,a_0^{-i}=a^{-i}]\bigg|\\
        &\le \gamma\|v_t^i+v_t^{-i}\|_\infty.
    \end{align*}
    In addition, we have
    \begin{align*}
    	\Big|\sum_{i=1,2}&\max_{\pi^i}(\pi^i)^\top \mathcal{T}^i(v_t^i)(s)\pi_{t,K}^{-i}(s)\Big| \\
        & = \left|\sum_{i=1,2} \max_{\pi^i} \left\{(\pi^i-\pi_{t,K}^i(s))^\top \mathcal{T}^i(v^i_t)(s) \pi_{t,K}^{-i}(s)\right\}\right|+\left|\sum_{i=1,2}(\pi_{t,K}^i(s))^\top \mathcal{T}^i(v^i_t)(s) \pi_{t,K}^{-i}(s)\right|\\
    	&\leq \sum_{i=1,2}\max_{\pi^i}\bigg\{(\pi^i-\pi_{t,K}^i(s))^\top \mathcal{T}^i(v^i_t)(s) \pi_{t,K}^{-i}(s) + \frac{1}{\eta} \cH(\pi^i)-\frac{1}{\eta} \cH(\pi_{t,K}^i(s))\bigg\}\\
    	&\quad+\left|\sum_{i=1,2}(\pi_{t,K}^i(s))^\top \mathcal{T}^i(v^i_t)(s) \pi_{t,K}^{-i}(s)\right| + \frac{8}{\eta} \sqrt{A_{\max}},
    \end{align*}
    then 
    \begin{align*}
        \left|\sum_{i=1,2}\max_{\pi^i}(\pi^i)^\top \mathcal{T}^i(v_t^i)(s)\pi_{t,K}^{-i}(s)\right| \le \: &V_{v_t,s}(\pi_{t,K}^i(s),\pi_{t,K}^{-i}(s))+\frac{8}{\eta} \sqrt{A_{\max}}\\
        & \quad +\left|\sum_{i=1,2}(\pi_{t,K}^i(s))^\top \mathcal{T}^i(v^i_t)(s) \pi_{t,K}^{-i}(s)\right| \\
    	\leq  \: &V_{v_t,s}(\pi_{t,K}^i(s),\pi_{t,K}^{-i}(s))+ \frac{8}{\eta} \sqrt{A_{\max}} \\
        & \quad +\max_{a^i,a^{-i}}\left|\mathcal{T}^i(v_t^i)(s,a^i,a^{-i})+\mathcal{T}^{-i}(v_t^{-i})(s,a^i,a^{-i})\right|\\
    	\leq  \: &V_{v_t,s}(\pi_{t,K}^i(s),\pi_{t,K}^{-i}(s))+ \frac{8}{\eta} \sqrt{A_{\max}} +\gamma\|v_t^i+v_t^{-i}\|_\infty.
    \end{align*}
    It follows that
    \begin{align}\label{eq:355}
    	\hat{E}_3\leq 2\gamma\|v_t^i+v_t^{-i}\|_\infty+\max_{s}V_{v_t,s}(\pi_{t,K}^i(s),\pi_{t,K}^{-i}(s))+\frac{8}{\eta} \sqrt{A_{\max}}.
    \end{align}

    Substituting the upper bounds we obtained for the terms $\{E_j\}_{1\leq j\leq 4}$ into Equation \ref{eq:connect1} 
    \begin{align*}
    	\|v^{-i}_*-v^{-i}_{\pi_{t,K}^{-i},*}\|_\infty\leq\;&\gamma\|v^i_{*,\pi_{t,K}^{-i}}-v^i_{*}\|_\infty+2\gamma\|v_t^i+v_t^{-i}\|_\infty+2\gamma\|v^i_t-v^i_{*}\|_\infty \\
        & +\max_{s}V_{v_t,s}(\pi_{t,K}^i(s),\pi_{t,K}^{-i}(s))+ \frac{8}{\eta} \sqrt{A_{\max}}\\
    	=\;&\gamma\|v^{-i}_*-v^{-i}_{\pi_{t,K}^{-i},*}\|_\infty+2\gamma\|v_t^i+v_t^{-i}\|_\infty+2\gamma\|v^i_t-v^i_{*}\|_\infty\\
    	& +\max_{s}V_{v_t,s}(\pi_{t,K}^i(s),\pi_{t,K}^{-i}(s))+ \frac{8}{\eta} \sqrt{A_{\max}},
    \end{align*}
    which implies
    \begin{align*}
    	\|v^{-i}_* \!-\! v^{-i}_{\pi_{t,K}^{-i},*}\|_\infty \!\leq\! & \frac{1}{1-\gamma}\bigg(\!2\|v_t^i+v_t^{-i}\|_\infty \!+\! 2\|v^i_t-v^i_{*}\|_\infty \!+\! \max_{s}V_{v_t,s}(\pi_{t,K}^i(s),\pi_{t,K}^{-i}(s)) \!+\! \frac{8}{\eta} \sqrt{A_{\max}}\!\bigg).
    \end{align*}
    Similarly, we also have
    \begin{align*}
    	\|v^i_*\!-\!v^i_{\pi_{t,K}^i,*}\|_\infty \!\leq\! \frac{1}{1-\gamma}\bigg(\!2\|v_t^i+v_t^{-i}\|_\infty \!+\! 2\|v^i_t-v^i_{*}\|_\infty \!+\! \max_{s}V_{v_t,s}(\pi_{t,K}^i(s),\pi_{t,K}^{-i}(s)) \!+\! \frac{8}{\eta} \sqrt{A_{\max}}\!\bigg).
    \end{align*}
    Substituting the previous two inequalities into Equation \ref{eq:last_policy_bound} we get
    \begin{equation*}
    	\|v^i_{*,\pi_{t,K}^{-i}}\!\!-v^i_{\pi_{t,K}^i,\pi_{t,K}^{-i}}\!\|_\infty \!\leq\! \frac{2}{1-\gamma}\bigg(\!2\|v_t^i+v_t^{-i}\|_\infty\!\!+\!2\|v^i_t-v^i_{*}\|_\infty \!\!+\! \max_{s}V_{v_t,s}(\pi_{t,K}^i(s),\pi_{t,K}^{-i}(s)) \!+\!\! \frac{8}{\eta} \sqrt{A_{\max}}\bigg).
    \end{equation*}
\end{proof}

At this point, it remains to bound the first two terms on the r.h.s in lemma \ref{lem:Nash_Gap_Decomp}. The next two lemmas achieve this purpose by presenting one-step drift inequalities for the relevant terms.
\begin{lemma}\label{lem:outer-loop}
	It holds for all $t\geq 0$ and $i=1,2$ that
	\begin{align*}
		\|v_{t+1}^i-v_*^i\|_\infty
		\leq \;&\gamma  \|v_t^i-v_*^i\|_\infty+2\max_{s\in\mathcal{S}}V_{v_t,s}(\pi_{t,K}^i(s),\pi_{t,K}^{-i}(s)) + \frac{16}{\eta} \sqrt{A_{\max}}\nonumber\\
		&+\max_{s\in\mathcal{S}}\|\mathcal{T}^i(v^i_t)(s) \pi_{t,K}^{-i}(s)- q_{t,K}^i(s)\|_\infty+2\gamma\|v_t^i+v_t^{-i}\|_\infty.
	\end{align*}
\end{lemma}
\begin{proof}
For any $i\in \{1,2\}$, we have by the outer-loop update equation (cf. Line \ref{line:VIteration}) of Algorithm \ref{algorithm:TBRVI} that
\begin{align*}
	v_{t+1}^i(s)= \pi_{t,K}^i(s)^\top q_{t,K}^i(s) = \textit{val}^i(\mathcal{T}^i(v^i_t)(s))+\pi_{t,K}^i(s)^\top q_{t,K}^i(s)-\textit{val}^i(\mathcal{T}^i(v^i_t)(s))
\end{align*}
Since $\textit{val}^i(\mathcal{T}^i(v_*^i)(s))=\mathcal{B}^i(v_*^i)(s)=v_*^i(s)$, we have
\begin{align}
	|v_{t+1}^i(s)-v_*^i(s)|
	=|\textit{val}^i(\mathcal{T}^i(v^i_t)(s))-\textit{val}^i(\mathcal{T}^i(v_*^i)(s))| + |\pi_{t,K}^i(s)^\top q_{t,K}^i(s)-\textit{val}^i(\mathcal{T}^i(v^i_t)(s))|.\label{eq1:prop:outer}
\end{align}
For the first term on the r.h.s of Equation \ref{eq1:prop:outer}, we have by the contraction property of the minimax Bellman operator that
\begin{align*}
	\left|\textit{val}^i(\mathcal{T}^i(v^i_t)(s))-\textit{val}^i(\mathcal{T}^i(v_*^i)(s))\right|=\;&\left|\mathcal{B}^i(v_t^i)(s)-\mathcal{B}^i(v_*^i)(s)\right|\\
	\leq\;& \gamma \|v^i_t-v_*^i\|_\infty.
\end{align*}
For the second term on the r.h.s of Equation \ref{eq1:prop:outer}, we have
\begin{align*}
	\left|\pi_{t,K}^i(s)^\top q_{t,K}^i(s)-\textit{val}^i(\mathcal{T}^i(v^i_t)(s))\right|\leq \;&\underbrace{\left|\max_{\pi^i}(\pi^i)^\top \mathcal{T}^i(v^i_t)(s) \pi_{t,K}^{-i}(s)-\pi_{t,K}^i(s)^\top q_{t,K}^i(s)\right|}_{T_1}\\
	&+\underbrace{\left|\max_{\pi^i}(\pi^i)^\top \mathcal{T}^i(v^i_t)(s) \pi_{t,K}^{-i}(s)-\textit{val}^i(\mathcal{T}^i(v^i_t)(s))\right|}_{T_2}
\end{align*}
For the term $T_1$, we have
\begin{align*}
	T_1
	\leq \;&\left|\max_{\pi^i}(\pi^i)^\top \mathcal{T}^i(v^i_t)(s) \pi_{t,K}^{-i}(s)-(\pi_{t,K}^i(s))^\top \mathcal{T}^i(v^i_t)(s) \pi_{t,K}^{-i}(s)\right|\\
	&+\left|(\pi_{t,K}^i(s))^\top \mathcal{T}^i(v^i_t)(s) \pi_{t,K}^{-i}(s)-\pi_{t,K}^i(s)^\top q_{t,K}^i(s)\right|\\
	\leq \;&\max_{\pi^i}(\pi^i)^\top \mathcal{T}^i(v^i_t)(s) \pi_{t,K}^{-i}(s)-(\pi_{t,K}^i(s))^\top \mathcal{T}^i(v^i_t)(s) \pi_{t,K}^{-i}(s)\\
	&+\|\mathcal{T}^i(v^i_t)(s) \pi_{t,K}^{-i}(s)- q_{t,K}^i(s)\|_\infty\\
	\leq \;&\sum_{i=1,2}\left\{\max_{\pi^i}(\pi^i-\pi_{t,K}^i(s))^\top \mathcal{T}^i(v^i_t)(s) \pi_{t,K}^{-i}(s)\right\}\\
	&+\|\mathcal{T}^i(v^i_t)(s) \pi_{t,K}^{-i}(s)- q_{t,K}^i(s)\|_\infty\\
	\leq \;&\sum_{i=1,2}\left\{\max_{\pi^i}(\pi^i-\pi_{t,K}^i(s))^\top \mathcal{T}^i(v^i_t)(s) \pi_{t,K}^{-i}(s)+\frac{1}{\eta} \nu(\pi^i)- \frac{1}{\eta} \nu(\pi_{t,K}^i(s))\right\}\\
	&+\frac{8}{\eta} \sqrt{A_{\max}}+\|\mathcal{T}^i(v^i_t)(s) \pi_{t,K}^{-i}(s)- q_{t,K}^i(s)\|_\infty\\
	\leq \;&V_{v_t,s}(\pi_{t,K}^i(s),\pi_{t,K}^{-i}(s))+\frac{8}{\eta} \sqrt{A_{\max}}+\|\mathcal{T}^i(v^i_t)(s) \pi_{t,K}^{-i}(s)- q_{t,K}^i(s)\|_\infty.
\end{align*}
Note that $T_2$ is exactly the term $\hat{E}_3$ we analyzed in proving Lemma \ref{lem:Nash_Gap_Decomp}. Therefore, we have from Equation \ref{eq:355} that
\begin{align*}
	T_2\leq 2\gamma\|v_t^i+v_t^{-i}\|_\infty+\max_{s}V_{v_t,s}(\pi_{t,K}^i(s),\pi_{t,K}^{-i}(s))+\frac{8}{\eta} \sqrt{A_{\max}}.
\end{align*}
It follows that
\begin{align*}
	\left|\pi_{t,K}^i(s)^\top q_{t,K}^i(s)-\textit{val}^i(\mathcal{T}^i(v^i_t)(s))\right| \leq &T_1+T_2\\
	\leq & 2\max_{s}V(\pi_{t,K}^i(s),\pi_{t,K}^{-i}(s))+2\gamma\|v_t^i+v_t^{-i}\|_\infty\\
	&+\max_{s}\|\mathcal{T}^i(v^i_t)(s) \pi_{t,K}^{-i}(s)- q_{t,K}^i(s)\|_\infty+\frac{16}{\eta} \sqrt{A_{\max}}.
\end{align*}
Using the upper bounds we obtained for the two terms on the r.h.s of equation (\ref{eq1:prop:outer})
\begin{align*}
	|v_{t+1}^i(s)-v_*^i(s)|
	\leq \;&\gamma  \|v_t^i-v_*^i\|_\infty+2\max_{s\in\mathcal{S}}V(\pi_{t,K}^i(s),\pi_{t,K}^{-i}(s))+\frac{16}{\eta} \sqrt{A_{\max}}\\
	&+\max_{s\in\mathcal{S}}\|\mathcal{T}^i(v^i_t)(s) \pi_{t,K}^{-i}(s)- q_{t,K}^i(s)\|_\infty+2\gamma\|v_t^i+v_t^{-i}\|_\infty.
\end{align*}
Since the r.h.s of the previous inequality does not depend on $s$, we have for any $i\in \{1,2\}$ that
\begin{align*}
	\|v_{t+1}^i-v_*^i\|_\infty
	\leq \;&\gamma  \|v_t^i-v_*^i\|_\infty+2\max_{s\in\mathcal{S}}V(\pi_{t,K}^i(s),\pi_{t,K}^{-i}(s))+\frac{16}{\eta} \sqrt{A_{\max}}\\
	&+\max_{s\in\mathcal{S}}\|\mathcal{T}^i(v^i_t)(s) \pi_{t,K}^{-i}(s)- q_{t,K}^i(s)\|_\infty+2\gamma\|v_t^i+v_t^{-i}\|_\infty.
\end{align*}    
\end{proof}

\begin{lemma}[Lemma A.6 of \cite{chen2023finite}]\label{lem:outer-sum}
	It holds for all $t\geq 0$ that
	\begin{align*}
		\|v_{t+1}^i+v_{t+1}^{-i}\|_\infty\leq \gamma\|v_t^i+v_t^{-i}\|_\infty+\sum_{i=1,2}\max_{s\in\mathcal{S}}\| q_{t,K}^i(s)-\mathcal{T}^i(v_t^i)(s)\pi_{t,K}^{-i}(s)\|_\infty.
	\end{align*}
\end{lemma}


\subsection{Inner-loop}

\subsubsection{Analyzing the strategy update}

For readability purposes, and in this section only, we denote $X^i = \mathcal{T}^i(v)(s,.,.) \in \bR^{|\cA^i|\times|\cA^{-i}|}$ and we define
\begin{equation*}
    V_X \left(\pi^i, \pi^{-i}\right)=\sum_{i=1,2} \max _{\hat{\pi}^i \in \Delta^{\left|\mathcal{A}^i\right|}}\left\{\left(\hat{\pi}^i-\pi^i\right)^{\top} X_i \pi^{-i}+\frac{1}{\eta} \cH\left(\hat{\pi}^i\right)-\frac{1}{\eta}  \cH\left(\pi^i\right)\right\}
\end{equation*}
this is introduced to simplify the notation of $V_{v, s} (\pi^i, \pi^{-i})$ defined in Equation \ref{eq:Lyapunov_pi}. Indeed, it is easy to see that
 $V_{v, s}(\cdot, \cdot) \eqdef V_{\big(\mathcal{T}^i\left(v^i\right)(s), \mathcal{T}^{-i}\left(v^{-i}\right)(s)\big)}(\cdot, \cdot)$.

We use $\nabla_1V_X(\cdot,\cdot)$ (respectively, $\nabla_2V_X(\cdot,\cdot)$) to denote the gradient concerning the first (respectively, second) argument. The following lemma establishes the strong convexity and the smoothness of $V_X(\pi^i,\pi^{-i})$.

\begin{lemma}\label{lem:properties_Lyapunov}
	The function $V_X(\cdot,\cdot)$ has the following properties.
	\begin{enumerate}
		\item For all $\pi^{-i}\in\Delta^{|\mathcal{A}^{-i}|}$, $V_X(\pi^i,\pi^{-i})$ as a function of $\pi^i$ is $1/\eta$ -- strongly convex with respect to $\|\cdot\|_2$.
		\item For any $\delta_i>0$ and $\pi^{-i}\in\Delta^{|\mathcal{A}^{-i}|}$, the function $V_X(\cdot,\pi^{-i})$ is $M_{\eta,i}$ -- smooth on $\{\pi^i\in\Delta^{|\mathcal{A}^i|}\mid \min_{a^i}\pi^i(a^i)\geq \delta_i\}$ with respect to $\|\cdot\|_2$, where $M_{\eta,i} = \left(2\sqrt{2} \:\eta  \:\sigma_{\max}^2(X_{-i}) \:|A_{\max}^{-i}| + \frac{1}{\eta \delta_i^{3/2}}\right)$.
		\item It holds for any $(\pi^i,\pi^{-i})$ that
		\begin{align*}
			\langle \nabla_1V_X(\pi^i,\pi^{-i}),\ts(X_i\pi^{-i})-\pi^i \rangle + &\langle \nabla_2V_X(\pi^i,\pi^{-i}),\ts(X_{-i}\pi^i)-\pi^{-i} \rangle\\
			\leq\;& -\frac{7}{8}V_X(\pi^i,\pi^{-i}) + 16 \eta \|X_i+X_{-i}^\top \|_2^2.
		\end{align*}
		\item For any $u^i\in\mathbb{R}^{|\mathcal{A}^i|},u^{-i}\in\mathbb{R}^{|\mathcal{A}^{-i}|}$ , we have for all $(\pi^i,\pi^{-i})\in \{\pi^i\in\Delta^{|\mathcal{A}^i|},\pi^{-i}\in\Delta^{|\mathcal{A}^{-i}|}\mid \min_{a^i}\pi^i(a^i)\geq \delta_i,\min_{a^{-i}}\pi^{-i}(a^{-i})\geq \delta_{-i}\}$ (where $\delta_i,\delta_{-i}>0$) that
		\begin{align*}
			\langle \nabla_1V_X(\pi^i,\pi^{-i}),&\ts(u^i)-\ts(X_i\pi^{-i})\rangle + \langle \nabla_2V_X(\pi^i,\pi^{-i}),\ts(u^{-i})-\ts(X_{-i}\pi^i)\rangle\\
			\leq &\left(\frac{1}{\eta \delta_{i}^{3/2}} + \frac{1}{\eta \delta_{-i}^{3/2}} + \|X_i\|_2+\|X_{-i}\|_2\right) \bigg[2\Bar{c} \eta V_X(\pi^i,\pi^{-i})\\
		&+ \frac{8\eta^2  A_{\max}^2}{\Bar{c}}\|u^i-X_i\pi^{-i} \|_2^2 + \frac{8\eta^2 A_{\max }^2}{\Bar{c}}\| u^{-i}-X_i\pi^i\|_2^2\bigg]
		\end{align*}
		where $\Bar{c}$ is any positive real number.
	\end{enumerate}
\end{lemma}


\begin{proof}
    First, observe that 
    \begin{equation*}
        \arg \max _{\hat{\pi}^i \in \Delta^{\left|\mathcal{A}^i\right|}}\left\{\left(\hat{\pi}^i\right)^{\top} X_i \pi^{-i}+\frac{1}{\eta} \cH\left(\hat{\pi}^i\right)\right\}=\ts_{\eta} \left(X_i \pi^{-i}\right) .
    \end{equation*}
    Therefore, the function $V_X(\cdot, \cdot)$ can be equivalently written as
    \begin{align*}
        V_X\left(\pi^i, \pi^{-i}\right)=\sum_{i=1,2}\Bigg[&\left(\ts_{\eta}\left(X_i \pi^{-i}\right)\right)^{\top} X_i \pi^{-i}+\frac{1}{\eta} \cH\left(\ts_{\eta}\left(X_i \pi^{-i}\right)\right) \\
        &-\left(\pi^i\right)^{\top} X_i \pi^{-i}-\frac{1}{\eta} \cH\left(\pi^i\right)\Bigg]
    \end{align*}
    
    \textit{1) Convexity:} first, it is easy to show that the negative Tsallis--entropy $-\cH(\cdot)$ is $1$--strongly convex with respect to $\|\|_2$. Indeed, $\nabla^2 (-\cH(\pi)) = \operatorname{diag}((1/\pi_a^{3/2})_{a \in \cA})$ then $\nabla^2 (-\cH(\pi)) \ge I_{\cA}$, \ie it is $1$--strongly convex.
        
    Second, we have that $$\left(\ts_\eta\left(X_{-i} \pi^i\right)\right)^{\top} X_{-i} \pi^i+\frac{1}{\eta} \cH\left(\ts\left(X_{-i} \pi^i\right)\right)=\max _{\hat{\pi}^{-i} \in \Delta^{\mid \mathcal{A}}-i \mid}\left\{\left(\hat{\pi}^{-i}\right)^{\top} X_{-i} \pi^i+\frac{1}{\eta} \cH\left(\hat{\pi}^{-i}\right)\right\},$$ 
    which is convex as a maximum of convex functions.
    
    Therefore, the function $V_X \left(\cdot, \pi^{-i}\right)$ is $1/\eta$ - strongly convex with respect to $\|\cdot\|_2$ uniformly for all $\pi^{-i}$.

    \textit{2) Smoothness:} first, from the Hessian $\nabla^2 (\cH(\pi)) = \operatorname{diag}((1/\pi_a^{3/2})_{a \in \cA})$, it is clear that the negative Tsallis entropy is $1/\delta_i^{3/2}$-- smooth on $\{\pi^i\in\Delta^{|\mathcal{A}^i|}\mid \min_{a^i}\pi^i(a^i)\geq \delta_i\}$ with respect to $\|\cdot\|_2$.

    Second, we have
    \begin{align*}
        \nabla_{\pi^i}\Big(\ts_\eta\left(X_{-i} \pi^i\right)^{\top} X_{-i} \pi^i+&\frac{1}{\eta}\cH\left(\ts_\eta\left(X_{-i} \pi^i\right)\right)\Big) \\
        &= \nabla_{\pi^i} \max _{\hat{\pi}^{-i} \in \Delta^{\mid \mathcal{A}}-i \mid}\left\{\left(\hat{\pi}^{-i}\right)^{\top} X_{-i} \pi^i+\frac{1}{\eta}\cH\left(\hat{\pi}^{-i}\right)\right\} \\
        &= X_{-i}^{\top} \ts_\eta\left(X_{-i} \pi^i\right) .
    \end{align*}
    where the first line follows using Danskin's theorem.
    Since there aren't readily usable formulas for the gradient of the Tsallis weights, we will use the standard characterization of smoothness. Let $\pi_1 , \pi_2 \in \Delta^{|\cA^i|}$, we have by lemma \ref{lem:lipschitz_tsallis} we have:
    \begin{align*}
        X_{-i}^{\top} \ts_\eta\left(X_{-i} \pi_1\right) - X_{-i}^{\top} \ts_\eta\left(X_{-i} \pi_2\right) &= X_{-i}^{\top} \left(\ts_\eta\left(X_{-i} \pi_1\right) - \ts_\eta\left(X_{-i} \pi_2\right)\right) \\
        &\le \sigma_{\max}(X_{-i}) \|\ts_\eta\left(X_{-i} \pi_1\right) - \ts_\eta\left(X_{-i} \pi_2 \right) \|_2 \\
        &\le \sigma_{\max}(X_{-i}) 2 \sqrt{2}\eta |\cA^{-i}| \|X_{-i} (\pi_1 - \pi_2) \|_2 \\
        &\le 2 \sqrt{2} \:\eta \: \sigma_{\max}^2 (X_{-i})\: |\cA^{-i}| \: \|\pi_1 - \pi_2 \|_2.
    \end{align*}
    
    Since we showed before that $$\nabla_{\pi^i}\left(\ts_\eta\left(X_{-i} \pi^i\right)^{\top} X_{-i} \pi^i+\frac{1}{\eta}\cH\left(\ts_\eta\left(X_{-i} \pi^i\right)\right)\right) = X_{-i}^{\top} \ts_\eta\left(X_{-i} \pi^i\right),$$
    we deduce that the function $\ts_\eta\left(X_{-i} \pi^i\right)^{\top} X_{-i} \pi^i+\frac{1}{\eta}\cH\left(\ts_\eta\left(X_{-i} \pi^i\right)\right)$ 
    is $2 \sqrt{2} \:\eta \: \sigma_{\max}(X_{-i})^2 \: |\cA^{-i}|$ smooth with respect to $\|\cdot\|_2$.

    Combining with the smoothness of Tsallis entropy, we conclude that $V_X \left(\cdot, \pi^{-i}\right)$ is $M_{\eta,i}$--smooth with respect to $\|\cdot\|_2$ uniformly for all $\pi^{-i}$, where we defined $M_{\eta,i} = 2\sqrt{2} \eta  \sigma_{\max}^2 X_{-i}) |\cA^{-i}| + \frac{1}{\eta \delta_i^{3/2}}.$

    \textit{3)} We first compute the gradient $\nabla_1 V_X(\pi^i,\pi^{-i})$ using Danskin's theorem:
	\begin{align}\label{eq:V_gradient}
    	\nabla_1 V_X(\pi^i,\pi^{-i}) = - (X_i+X_{-i}^\top)\pi^{-i}-\frac{1}{\eta} \nabla \cH(\pi^i)+X_{-i}^\top \ts(X_{-i} \pi^i).
	\end{align}
	It follows that
	\begin{align*}
		\langle \nabla_1V_X(\pi^i,\pi^{-i}),&\ts(X_i\pi^{-i})-\pi^i \rangle \\
        & = \langle -(X_i+X_{-i}^\top)\pi^{-i}-\frac{1}{\eta} \nabla \cH(\pi^i)+X_{-i}^\top \ts(X_{-i} \pi^i), \ts(X_i\pi^{-i})-\pi^i \rangle \\
		&= \langle -(X_i+X_{-i}^\top)\pi^{-i}-\frac{1}{\eta} \nabla \cH(\pi^i)+X_{-i}^\top \ts(X_{-i} \pi^i), \ts(X_i\pi^{-i})-\pi^i \rangle \\
		&\quad+\langle X_i\pi^{-i}+\frac{1}{\eta} \nabla \cH(\ts(X_i \pi^{-i})), \ts(X_i\pi^{-i})-\pi^i \rangle \numberthis \label{eq:oc}\\
		&=\frac{1}{\eta}\langle  \nabla \cH(\ts(X_i \pi^{-i}))- \nabla \cH(\pi^i), \ts(X_i\pi^{-i})-\pi^i \rangle \\
		&\quad+( \ts(X_{-i} \pi^i)-\pi^{-i})^\top X_{-i}( \ts(X_i\pi^{-i})-\pi^i ).
	\end{align*}
	where Equation \ref{eq:oc} from from the optimality condition $X_i\pi^{-i}+\frac{1}{\eta} \nabla \cH(\ts(X_i \pi^{-i}))=0$. 
 
    To proceed, observe that the concavity of $\cH(\cdot)$ and the optimality condition imply
	\begin{align*}
		\langle  \nabla \cH(\ts(X_i \pi^{-i})) &- \nabla \cH(\pi^i), \ts(X_i\pi^{-i})-\pi^i \rangle \\
         =\;&\langle   \nabla \cH(\pi^i)-\nabla \cH(\ts(X_i \pi^{-i})), \pi^i-\ts(X_i\pi^{-i}) \rangle\\
		=\;&\langle   \nabla \cH(\pi^i), \pi^i-\ts(X_i\pi^{-i}) \rangle-\langle\nabla \cH(\ts(X_i \pi^{-i})), \pi^i-\ts(X_i\pi^{-i}) \rangle\\
		\leq \;&\cH(\pi^i)-\cH(\ts(X_i\pi^{-i}))-\langle\nabla \cH(\ts(X_i \pi^{-i})), \pi^i-\ts(X_i\pi^{-i}) \rangle\\
		=\;&\cH(\pi^i)-\cH(\ts(X_i\pi^{-i}))+\eta\langle X_i\pi^{-i}, \pi^i-\ts(X_i\pi^{-i}) \rangle\\
		=\;&\eta \left[(\pi^i)^\top X_i \pi^{-i} +\frac{1}{\eta} \cH(\pi^i)-\max_{\hat{\pi}^i\in\Delta^{|\mathcal{A}^i|}}\left\{(\hat{\pi}^i)^\top X_i\pi^{-i}+ \frac{1}{\eta} \cH(\hat{\pi}^i)\right\}\right].
	\end{align*}
 
	Therefore, we have
	\begin{align*}
		\langle \nabla_1V_X(\pi^i,\pi^{-i}),\ts(X_i\pi^{-i})-\pi^i \rangle \leq \;&\Bigg[(\pi^i)^\top X_i \pi^{-i} + \frac{1}{\eta} \cH(\pi^i) \\
        & \quad-\max_{\hat{\pi}^i\in\Delta^{|\mathcal{A}^i|}}\left\{(\hat{\pi}^i)^\top X_i\pi^{-i}+ \frac{1}{\eta} \cH(\hat{\pi}^i)\right\}\Bigg]\\
		&\quad + ( \ts(X_{-i} \pi^i)-\pi^{-i})^\top X_{-i}( \ts(X_i\pi^{-i})-\pi^i )
	\end{align*}
 
	Similarly, we also have 
	\begin{align*}
		\langle \nabla_2V_X(\pi^i,\pi^{-i}),&\ts(X_{-i}\pi^i)-\pi^{-i} \rangle \\
        \leq \;&\left[(\pi^{-i})^\top X_{-i} \pi^i +\frac{1}{\eta} \cH(\pi^{-i})-\max_{\hat{\pi}^{-i}\in\Delta^{|\mathcal{A}^{-i}|}}\left\{(\hat{\pi}^{-i})^\top X_{-i}\pi^i+\frac{1}{\eta} \cH(\hat{\pi}^{-i})\right\}\right]\\
		&+( \ts(X_i \pi^{-i})-\pi^i)^\top X_i( \ts(X_{-i}\pi^i)-\pi^{-i} )
	\end{align*}
	Adding up the previous two inequalities we obtain
	\begin{align}
		\langle \nabla_1V_X(\pi^i,\pi^{-i})&,\ts(X_i\pi^{-i})-\pi^i \rangle + \langle \nabla_2V_X(\pi^i,\pi^{-i}),\ts(X_{-i}\pi^i)-\pi^{-i} \rangle\nonumber\\
		\leq \;&-V_X(\pi^i,\pi^{-i})+( \ts(X_i \pi^{-i})-\pi^i)^\top (X_i+X_{-i}^\top )( \ts(X_{-i}\pi^i)-\pi^{-i} ).\label{eq:gradient_V_1}
	\end{align}
 
	To control the second term on the r.h.s of Equation \ref{eq:gradient_V_1}, observe that
	\begin{align*}
		( \ts(X_i \pi^{-i})-\pi^i)^\top &(X_i+X_{-i}^\top )( \ts(X_{-i}\pi^i) -\pi^{-i}) \\
        \leq\;& \|\ts(X_i \pi^{-i})-\pi^i\|_2 \|X_i+X_{-i}^\top \|_2\| \ts(X_{-i}\pi^i)-\pi^{-i}\|_2 \\
		\leq\;& (\|\ts(X_i \pi^{-i})\|_2+\|\pi^i\|_2) \|X_i+X_{-i}^\top \|_2\| \ts(X_{-i}\pi^i)-\pi^{-i}\|_2 \\
		\leq \;&2\|X_i+X_{-i}^\top \|_2\| \ts(X_{-i}\pi^i)-\pi^{-i}\|_2 \\
		\leq \;&c_1\|X_i+X_{-i}^\top \|_2^2+\frac{1}{c_1}\| \ts(X_{-i}\pi^i)-\pi^{-i}\|_2^2\tag{This is true for all $c_1>0$} \\
		\leq \;&c_1\|X_i+X_{-i}^\top \|_2^2+\frac{1}{c_1}(\| \ts(X_{-i}\pi^i)-\pi^{-i}\|_2^2+\|\ts(X_i \pi^{-i})-\pi^i\|_2^2). \numberthis\label{eq:gradientV12}
	\end{align*}
 
	Next, note that the function
	\begin{align*}
	    F_{X_i}(\pi^i,\pi^{-i}):=\max_{\hat{\pi}^i}\left\{(\hat{\pi}^i-\pi^i)^\top X_i\pi^{-i} + \frac{1}{\eta} \cH(\hat{\pi}^i)-\frac{1}{\eta} \cH(\pi^i)\right\}
	\end{align*}
	is $\frac{1}{\eta}$-strongly convex as a function of $\pi^i$ uniformly for all $\pi^{-i}$. Therefore, we have
	\begin{align*}
	    F_{X_i}(\pi^i,\pi^{-i})=\;&F_{X_i}(\pi^i,\pi^{-i})-F_{X_i}(\ts(X_i\pi^{-i}),\pi^{-i})\\
	    =\;&F_{X_i}(\pi^i,\pi^{-i})-\min_{\pi^i}F_{X_i}(\pi^i,\pi^{-i})\\
	    \geq  \;&\frac{1}{2\eta} \|\ts(X_i\pi^{-i})-\pi^i\|_2^2,
	\end{align*}
	which is called the quadratic growth property in optimization literature. It follows that
	\begin{align*}
	    \|\ts(X_i\pi^{-i})-\pi^i\|_2^2\leq 2\eta \max_{\hat{\pi}^i}\left\{(\hat{\pi}^i-\pi^i)^\top X_i\pi^{-i}+\frac{1}{\eta} \cH(\hat{\pi}^i)-\frac{1}{\eta} \cH(\pi^i)\right\}.
	\end{align*}
	Similarly, we also have
	\begin{align*}
	    \|\ts(X_{-i}\pi^i)-\pi^{-i}\|_2^2\leq 2\eta \max_{\hat{\pi}^{-i}}\left\{(\hat{\pi}^{-i}-\pi^{-i})^\top X_{-i}\pi^i+\frac{1}{\eta} \cH(\hat{\pi}^{-i}) - \frac{1}{\eta} \cH(\pi^{-i})\right\}.
	\end{align*}
	Adding up the previous two inequalities
	\begin{align*}
	    \|\ts(X_{-i}\pi^i)-\pi^{-i}\|_2^2+\|\ts(X_i\pi^{-i})-\pi^i\|_2^2\leq 2\eta V_X(\pi^i,\pi^{-i}).
	\end{align*}
	Plugging the previous inequality in Equation \ref{eq:gradientV12}
	\begin{align*}
		(\ts(X_i \pi^{-i})-\pi^i)^\top & (X_i+X_{-i}^\top )( \ts(X_{-i}\pi^i) -\pi^{-i}) \\
		\leq \;&c_1\|X_i+X_{-i}^\top \|_2^2+\frac{1}{c_1}(\| \ts(X_{-i}\pi^i)-\pi^{-i}\|_2^2+\|\ts(X_i \pi^{-i})-\pi^i\|_2^2)\\
		\leq \;&c_1\|X_i+X_{-i}^\top \|_2^2+\frac{2\eta}{c_1}V_X(\pi^i,\pi^{-i})\\
		= \;& 16 \eta \|X_i+X_{-i}^\top \|_2^2+\frac{1}{8}V_X(\pi^i,\pi^{-i}),
	\end{align*}
	where the last line follows from choosing $c_1=16 \eta $. Using the previous inequality in Equation \ref{eq:gradient_V_1}
	\begin{align*}
		\langle \nabla_1V_X(\pi^i,\pi^{-i}),\ts(X_i\pi^{-i})-\pi^i \rangle + &\langle \nabla_2V_X(\pi^i,\pi^{-i}),\ts(X_{-i}\pi^i)-\pi^{-i} \rangle\\
		\leq \;&-\frac{7}{8}V_X(\pi^i,\pi^{-i}) + 16 \eta \|X_i+X_{-i}^\top \|_2^2.
	\end{align*}

    \textit{4)} For any $u^i\in\mathbb{R}^{|\mathcal{A}^i|}$, using the explicit expression of the gradient of $V_X(\cdot,\cdot)$ from Equation \ref{eq:V_gradient}
	\begin{align*}
		\langle \nabla_1 V_X(\pi^i,\pi^{-i}) &,\ts(u^i)-\ts(X_i\pi^{-i})\rangle \\
            = \;&\langle -(X_i+X_{-i}^\top)\pi^{-i}-\frac{1}{\eta} \nabla \cH(\pi^i)+X_{-i}^\top \ts(X_{-i} \pi^i),\ts(u^i)-\ts(X_i\pi^{-i})\rangle\\
		=\;&\langle -(X_i+X_{-i}^\top)\pi^{-i}-\frac{1}{\eta} \nabla \cH(\pi^i)+X_{-i}^\top \ts(X_{-i} \pi^i), \ts(u^i)-\ts(X_i\pi^{-i}) \rangle\\
		&+\langle X_i\pi^{-i}+\frac{1}{\eta} \nabla \cH(\ts(X_i \pi^{-i})), \ts(u^i)-\ts(X_i\pi^{-i}) \rangle\\
		=\;&\frac{1}{\eta}\langle  \nabla \cH(\ts(X_i \pi^{-i}))- \nabla \cH(\pi^i), \ts(u^i)-\ts(X_i\pi^{-i}) \rangle\\
		&+( \ts(X_{-i} \pi^i)-\pi^{-i})^\top X_{-i}( \ts(u^i)-\ts(X_i\pi^{-i}) )\\
		\leq \;&\frac{1}{\eta}\|\nabla \cH(\ts(X_i \pi^{-i}))- \nabla \cH(\pi^i)\|_2 \|\ts(u^i)-\ts(X_i\pi^{-i}) \|_2\\
		&+\|\ts(X_{-i} \pi^i)-\pi^{-i}\|_2 \|X_{-i}\|_2\| \ts(u^i)-\ts(X_i\pi^{-i})\|_2 \\
		\leq \;&\frac{1}{\eta \delta_i^{3/2}}\|\ts(X_i \pi^{-i})- \pi^i\|_2 \|\ts(u^i)-\ts(X_i\pi^{-i}) \|_2\\
		&+ \|X_{-i}\|_2\|\ts(X_{-i} \pi^i)-\pi^{-i}\|_2\| \ts(u^i)-\ts(X_i\pi^{-i})\|_2,
	\end{align*}
	where the last inequality follows from the smoothness of $\cH(\cdot)$ in Lemma.~\ref{lem:properties_Lyapunov} (2).
	Similarly, we also have for any $u^{-i}\in\mathbb{R}^{|\mathcal{A}^{-i}|}$ that 
	\begin{align*}
		\langle \nabla_2V_X(\pi^i,\pi^{-i}) &,\ts(u^{-i})-\ts(X_{-i}\pi^i) \rangle \\
            \leq & \frac{1}{\eta \delta_{-i}^{3/2}} \|\ts(X_{-i} \pi^i)- \pi^{-i}\|_2 \|\ts(u^{-i})-\ts(X_{-i}\pi^i) \|_2\\
		&+ \|X_i\|_2\|\ts(X_i \pi^{-i})-\pi^i\|_2\| \ts(u^{-i})-\ts(X_i\pi^i)\|_2.
	\end{align*}
	Adding up the previous two inequalities
	\begin{align*}
		\langle \nabla_1V_X(\pi^i,\pi^{-i}),& \ts(u^i)-\ts(X_i\pi^{-i})\rangle+\langle \nabla_2V_X(\pi^i,\pi^{-i}),\ts(u^{-i})-\ts(X_{-i}\pi^i)\rangle\\
		\leq \;&\left(\frac{1}{\eta \delta_{i}^{3/2}}+\frac{1}{\eta \delta_{-i}^{3/2}} + \|X_i\|_2+\|X_{-i}\|_2\right) \bigg(\sum_i \|\ts(X_i \pi^{-i})- \pi^i\|_2\bigg)\\
		&\times \left(\|\ts(u^i)-\ts(X_i\pi^{-i}) \|_2+\| \ts(u^{-i})-\ts(X_i\pi^i)\|_2\right)\\
		\leq \;&\frac{1}{2} \left(\frac{1}{\eta \delta_{i}^{3/2}}+\frac{1}{\eta \delta_{-i}^{3/2}} + \|X_i\|_2+\|X_{-i}\|_2\right) \\
        & \times \bigg[\Bar{c}\left(\|\ts(X_i \pi^{-i})- \pi^i\|_2+\|\ts(X_{-i} \pi^i)-\pi^{-i}\|_2\right)^2\\
		&\qquad +\frac{1}{\Bar{c}}\left(\|\ts(u^i)-\ts(X_i\pi^{-i}) \|_2+\| \ts(u^{-i})-\ts(X_i\pi^i)\|_2\right)^2\bigg]
	\end{align*}
        the last line is true for all $\Bar{c}>0$, then,
        \begin{align*}
            \langle \nabla_1V_X(\pi^i,\pi^{-i}),& \ts(u^i)-\ts(X_i\pi^{-i})\rangle+\langle \nabla_2V_X(\pi^i,\pi^{-i}),\ts(u^{-i})-\ts(X_{-i}\pi^i)\rangle\\
		\leq \;& \left(\frac{1}{\eta \delta_{i}^{3/2}}+\frac{1}{\eta \delta_{-i}^{3/2}} + \|X_i\|_2+\|X_{-i}\|_2\right) \Bar{c} \bigg( \sum_i \|\ts(X_i \pi^{-i})- \pi^i\|_2^2\bigg)\\
		&+\frac{1}{\Bar{c}}\|\ts(u^i)-\ts(X_i\pi^{-i}) \|_2^2+\frac{1}{\Bar{c}}\| \ts(u^{-i})-\ts(X_i\pi^i)\|_2^2\bigg] \\
		\leq \;& \left(\frac{1}{\eta \delta_{i}^{3/2}} + \frac{1}{\eta \delta_{-i}^{3/2}} + \|X_i\|_2+\|X_{-i}\|_2\right) \bigg[2\Bar{c} \eta V_X(\pi^i,\pi^{-i})\\
		&\quad +\frac{8\eta^2  A_{\max}^2}{\Bar{c}}\|u^i-X_i\pi^{-i} \|_2^2 +\frac{8\eta^2 A_{\max }^2}{\Bar{c}}\| u^{-i}-X_i\pi^i\|_2^2\bigg],
        \end{align*}
	where the penultimate line holds because $(a+b)^2\leq 2(a^2+b^2)$ for all $a,b\in\mathbb{R}$, and last line follows from the quadratic growth property of strongly convex functions and the Lipschitz continuity of the Tsallis function (\cf lemma \ref{lem:lipschitz_tsallis}).
    
\end{proof}

With the properties of $V_X(\cdot,\cdot)$ established above, we can now use it to study the strategy update $\pi_k^i$ and $\pi_k^{-i}$. Specifically, recall that $V_X$ is just a temporary notation in this section for $V_{v,s}$. Therefore, the newly proved smoothness of $V_{v,s}$, the update equation, and lemma \ref{lem:properties_Lyapunov} provide us with the drift inequality for $\mathcal{D}_\pi$ in the lemma below.

\begin{lemma}[Policy drift]\label{lem:policy_drift}
The following inequality holds for all $k\geq 0$:
\begin{align*}
	\sum_{s}\mathbb{E}[V_{v,s}(\pi_{k+1}^i(s),\pi_{k+1}^{-i}(s))]
	\leq \;&\left(1-\frac{3\beta_k}{4}\right)\sum_{s}\mathbb{E}[V_{v,s}(\pi_k^i(s),\pi_k^{-i}(s))]+\frac{4|\mathcal{S}|A_{\max }^2}{\ell_{\eta}(1-\gamma)^2}\beta_k^2\\
	&+\frac{2048A_{\max}^4\beta_k \eta^3}{\ell_{\eta}^3 (1-\gamma)^2}\sum_{i=1,2}\sum_{s}\mathbb{E}[\|q_k^i(s)-\mathcal{T}^i(v^i)(s)\pi_k^{-i}(s) \|_2^2]\\
	&+ 16|\mathcal{S}|A_{\max } \beta_k \eta \|v^i+v^{-i}\|_\infty^2.
\end{align*}
\end{lemma}

\begin{proof}
    Since $\min_{i=1,2}\min_{s,a^i}\pi_k^i(a^i|s)\geq \ell_\eta$ (by lemma \ref{lem:margins}), then lemma \ref{lem:properties_Lyapunov} (2) implies that the function $V_{v,s}(\pi^i,\pi^{-i})$ as a function of $\pi^i$ is $M_{\eta,i}$ -- smooth on $\{\pi^i\in\Delta^{|\mathcal{A}^i|}\mid \min_{a^i}\pi^i(a^i)\geq \ell_\eta\}$ uniformly for all $\pi^{-i}$, where
    \begin{align*}
        M_{\eta,i} \eqdef \left(2\sqrt{2} \:\eta  \:\sigma_{\max}^2(\mathcal{T}^{-i}(v^{-i})(s)) \:A_{\max } + \frac{1}{\eta \ell_eta^{3/2}}\right).
    \end{align*}
    We next bound $M_{\eta, i}$ from above.
    Since $\|v^i\|_\infty\leq 1/(1-\gamma)$ and $\|v^{-i}\|_\infty\leq 1/(1-\gamma)$, we have for any $(s,a^i,a^{-i})$ that
    \begin{align*}
        |\mathcal{T}^{-i}(v^{-i})(s,a^{-i},a^i)|\leq \;&|\mathcal{R}^{-i}(s,a^{-i},a^i)|+\gamma \mathbb{E}[|v^{-i}(S_1)|\mid S_0=s,A_0^i=a^i,A_0^{-i}=a^{-i}]\\
        \leq \;&1+\frac{\gamma}{1-\gamma} = \frac{1}{1-\gamma},
    \end{align*}
    which implies that
    \begin{align}\label{eqeq:cici}
        \sigma_{\max}(\mathcal{T}^{-i}(v^{-i})(s))=\;&\|\mathcal{T}^{-i}(v^{-i})(s))\|_2\leq \frac{\sqrt{|\mathcal{A}^i||\mathcal{A}^{-i}|}}{1-\gamma}\leq \frac{A_{\max }}{1-\gamma}.
    \end{align}
    As a result, we have by $\frac{1}{\eta}\leq 1$ and $\ell_\eta\leq 1$ that
    \begin{align*}
        M_{\eta,i}= \eta \sigma_{\max}^2(\mathcal{T}^{-i}(v^{-i})(s)) +\frac{1}{\eta \ell_{\eta}}\leq \frac{\eta A_{\max}^2}{(1-\gamma)^2}+\frac{1}{\eta \ell_{\eta}}\leq \frac{2A_{\max }^2}{\ell_{\eta}(1-\gamma)^2}:=M_\eta.
    \end{align*}
    Similarly, $V_{v,s}(\pi^i,\pi^{-i})$ is also $M_\eta$ -- smooth on the set $\{\pi^{-i}\in\Delta^{|\mathcal{A}^{-i}|}\mid \min_{a^{-i}}\pi^i(a^{-i})\geq \ell_\eta\}$ uniformly for all $\pi^i$.
    
    Using the smoothness of $V_{v,s}(\cdot,\cdot)$ established above, for any $s\in\mathcal{S}$, we have by the strategy update equation that  
    \begin{align*}
    	V_{v,s}(\pi_{k+1}^i(s),&\pi_{k+1}^{-i}(s)) \\
        \leq \;&V_{v,s}(\pi_k^i(s),\pi_k^{-i}(s))+\beta_k\langle \nabla_2V_{v,s}(\pi_k^i(s),\pi_k^{-i}(s)),\ts(q_k^{-i}(s))-\pi_k^{-i}(s) \rangle \\
    	&+\beta_k\langle \nabla_1V_{v,s}(\pi_k^i(s),\pi_{k+1}^{-i}(s)),\ts(q_k^i(s))-\pi_k^i(s) \rangle \\
    	&+\frac{M_\eta\beta_k^2}{2}\|\ts(q_k^i(s))-\pi_k^i(s)\|_2^2+\frac{M_\eta\beta_k^2}{2}\|\ts(q_k^{-i}(s))-\pi_k^{-i}(s)\|_2^2 \\
    	\leq \;&V_{v,s}(\pi_k^i(s),\pi_k^{-i}(s))\\
            &+\underbrace{\beta_k\langle \nabla_2V_{v,s}(\pi_k^i(s),\pi_k^{-i}(s)),\ts(\mathcal{T}^{-i}(v^{-i})(s)\pi_k^i(s))-\pi_k^{-i}(s) \rangle}_{\hat{N}_1} \\
    	&+\underbrace{\beta_k\langle \nabla_1V_{v,s}(\pi_k^i(s),\pi_{k+1}^{-i}(s)),\ts(\mathcal{T}^i(v^i)(s)\pi_k^{-i}(s))-\pi_k^i(s) \rangle}_{\hat{N}_2} \\
    	&+\underbrace{\beta_k\langle \nabla_2V_{v,s}(\pi_k^i(s),\pi_k^{-i}(s)),\ts(q_k^{-i}(s))-\ts(\mathcal{T}^{-i}(v^{-i})(s)\pi_k^i(s)) \rangle}_{\hat{N}_3} \\
    	&+\underbrace{\beta_k\langle \nabla_1V_{v,s}(\pi_k^i(s),\pi_{k+1}^{-i}(s)),\ts(q_k^i(s))-\ts(\mathcal{T}^i(v^i)(s)\pi_k^{-i}(s)) \rangle}_{\hat{N}_4} \\
    	&+2M_\eta\beta_k^2. \numberthis \label{eq:policy_overall}
    \end{align*}
    We next bound the terms $\{\hat{N}_j\}_{1\leq j\leq 4}$ on the r.h.s of Equation \ref{eq:policy_overall} using Lemma.~\ref{lem:properties_Lyapunov} (3) and (4). 
    
    First consider $\hat{N}_1+\hat{N}_2$. We have by Lemma.~\ref{lem:properties_Lyapunov} (3) that
    \begin{align*}
    	\hat{N}_1+\hat{N}_2
    	\leq \;&-\frac{7\beta_k}{8} V_{v,s}(\pi_k^i(s),\pi_k^{-i}(s))
    	+16 \beta_k \eta \| \mathcal{T}^i(v^i)(s)+\mathcal{T}^{-i}(v^{-i}(s)^\top \|_2^2.
    \end{align*}
    To proceed, note that for any $\pi^{-i}\in \mathbb{R}^{|\mathcal{A}^{-i}|}$ satisfying $\|\pi^{-i}\|_2=1$, we have
    \begin{align*}
    	\| (\mathcal{T}^i(v^i)(s)+&\mathcal{T}^{-i}(v^{-i}(s)^\top)\pi^i \|_2^2 \\
        = &\gamma^2 \sum_{a^i}\left[\sum_{a^{-i}} \mathbb{E}[v^i(s_1)+v^{-i}(s_1)\mid s_0=s,a_0^i=a^i,a_0^{-i}=a^{-i}]\pi^{-i}(a^{-i})\right]^2\\
    	\leq \;&\gamma^2\|v^i+v^{-i}\|_\infty^2 \sum_{a^i}\left[\sum_{a^{-i}} \pi^{-i}(a^{-i})\right]^2\\
    	\leq \;&\gamma^2A_{\max}\|v^i+v^{-i}\|_\infty^2,
    \end{align*}
    which implies 
    \begin{align*}
    	\| \mathcal{T}^i(v^i)(s)+\mathcal{T}^{-i}(v^{-i}(s)^\top \|_2^2\leq \gamma^2A_{\max }\|v^i+v^{-i}\|_\infty^2.
    \end{align*}
    It follows that
    \begin{align*}
    	\hat{N}_1+\hat{N}_2
    	\leq \;&-\frac{7\beta_k}{8} V_{v,s}(\pi_k^i(s),\pi_k^{-i}(s))
    	+16 A_{\max } \beta_k \eta \|v^i+v^{-i}\|_\infty^2.
    \end{align*}

    We next consider $\hat{N}_3+\hat{N}_4$. Since
    \begin{align*}
    	\max(\|\mathcal{T}^i(v^i)(s)\|_2 ,\|\mathcal{T}^{-i}(v^{-i})(s)\|_2)\leq \frac{ A_{\max }}{1-\gamma},\tag{See Equation \ref{eqeq:cici}}
    \end{align*}
    we have by Lemma.~\ref{lem:properties_Lyapunov} (4) that
    \begin{align*}
    	\hat{N}_3+\hat{N}_4 \leq\;& 2\beta_k\left(\frac{1}{\eta \ell_{\eta}^{3/2}}+\frac{A_{\max}}{1-\gamma}\right)\bigg[2\Bar{c}\eta V_{v,s}(\pi_k^i(s),\pi_k^{-i}(s))\\
    	&+\frac{8\eta^2 A_{\max}^2}{\Bar{c}}\|q_k^i(s)-\mathcal{T}^i(v^i)(s)\pi_k^{-i}(s) \|_2^2+ \frac{8\eta^2  A_{\max}^2}{\Bar{c}} \| q_k^{-i}(s)-\mathcal{T}^{-i}(v^{-i})(s)\pi_k^i(s)\|_2^2\bigg]
    \end{align*}
    for any $\Bar{c}>0$.
    By choosing $\Bar{c}=\frac{1}{32 \eta }(\frac{1}{\eta \ell_{\eta}^{3/2}}+\frac{ A_{\max}}{1-\gamma})^{-1}$,
    we have that
    \begin{align*}
    	\hat{N}_3+\hat{N}_4
    	\leq\;& \frac{\beta_k}{8} V_{v,s}(\pi_k^i(s),\pi_k^{-i}(s)) \\
            & + 512 \beta_k\left(\frac{1}{\eta \ell_{\eta}^{3/2}}+\frac{ A_{\max}}{1-\gamma}\right)^2 \eta^3  A_{\max}^2 \|q_k^i(s)-\mathcal{T}^i(v^i)(s)\pi_k^{-i}(s) \|_2^2\\
    	&+ 512 \beta_k\left(\frac{1}{\eta \ell_{\eta}^{3/2}}+\frac{ A_{\max}}{1-\gamma}\right)^2 \eta^3  A_{\max}^2 \| q_k^{-i}(s)-\mathcal{T}^{-i}(v^{-i})(s)\pi_k^i(s)\|_2^2\\
    	\leq\;& \frac{\beta_k}{8} V_{v,s}(\pi_k^i(s),\pi_k^{-i}(s))+\frac{2048  A_{\max}^4 \beta_k \eta^3}{\ell_{\eta}^3 (1-\gamma)^2}\|q_k^i(s)-\mathcal{T}^i(v^i)(s)\pi_k^{-i}(s) \|_2^2\\
    	&+\frac{2048  A_{\max}^4 \beta_k \eta^3}{\ell_{\eta}^3 (1-\gamma)^2} \| q_k^{-i}(s)-\mathcal{T}^{-i}(v^{-i})(s)\pi_k^i(s)\|_2^2.
    \end{align*}
    
    Finally, using the upper bounds we obtained for the terms $\hat{N}_1+\hat{N}_2$ and $\hat{N}_3+\hat{N}_4$ in Equation \ref{eq:policy_overall}
    \begin{align*}
    	V_{v,s}(\pi_{k+1}^i(s),\pi_{k+1}^{-i}(s)) \leq \;&\left(1-\frac{3\beta_k}{4}\right)V_{v,s}(\pi_k^i(s),\pi_k^{-i}(s))+ 16 A_{\max} \beta_k \eta \|v^i+v^{-i}\|_\infty^2.\\
    	&+ \frac{2048  A_{\max}^4 \beta_k \eta^3}{\ell_{\eta}^3 (1-\gamma)^2}\|q_k^i(s)-\mathcal{T}^i(v^i)(s)\pi_k^{-i}(s) \|_2^2\\
    	&+ \frac{2048  A_{\max}^4 \beta_k \eta^3}{\ell_{\eta}^3 (1-\gamma)^2} \| q_k^{-i}(s)-\mathcal{T}^{-i}(v^{-i})(s)\pi_k^i(s)\|_2^2\\
    	&+\frac{4 A_{\max}^2}{\ell_{\eta}(1-\gamma)^2}\beta_k^2.
    \end{align*}
    Summing up both sides of the previous inequality for all $s$ and then taking expectation, we deduce the desired result.
\end{proof}

\subsubsection{Analyzing the q-function update}

Consider $q_k^i$ generated by the algorithm. We begin by reformulating the update of the $q$-function as a stochastic approximation algorithm for estimating a time-varying target. Let $F^i:\mathbb{R}^{|\mathcal{S}| |\mathcal{A}^i|}\times \mathcal{S}\times \mathcal{A}^i\times \mathcal{A}^{-i}\times \mathcal{S}\mapsto \mathbb{R}^{|\mathcal{S}| |\mathcal{A}^i|}$ be an operator defined as
\begin{align*}
	[F^i(q^i,s_0,a_0^i,a_0^{-i},s_1)](s,a^i)=\mathds{1}_{\{(s,a^i)=(s_0,a_0^i)\}}\left(\mathcal{R}^i(s_0,a_0^i,a_0^{-i})+\gamma v^i(s_1)-q^i(s_0,a_0^i)\right)
\end{align*}
for all $(q^i,s_0,a_0^i,a_0^{-i},s_1)$ and $(s,a^i)$. Then the $q$--function update can be compactly written as
\begin{align}\label{sa:reformulation}
	q_{k+1}^i=q_k^i+\alpha_k F^i(q_k^i,S_k,A_k^i,A_k^{-i},S_{k+1}).
\end{align}
Denote the stationary distribution of the Markov chain $\{S_k\}$ induced by the joint strategy $\pi_k=(\pi_k^i,\pi_k^{-i})$ by $\pi_k\in\Delta^{|\mathcal{S}|}$, the existence and uniqueness of which is guaranteed by Lemma \ref{lem:margins} and Lemma \ref{lem:zhang}.
Let $\Bar{F}_k^i:\mathbb{R}^{|\mathcal{S}| |\mathcal{A}^i|}\mapsto \mathbb{R}^{|\mathcal{S}| |\mathcal{A}^i|}$ be defined as
\begin{align*}
	\Bar{F}_k^i(q^i)=\mathbb{E}_{s_0\sim \pi_k(\cdot),a_0^i\sim \pi_k^i(\cdot|s_0), a_k^{-i}\sim \pi_k^{-i}(\cdot|s_0), S_1\sim P(\cdot|s_0,a_0^i,a_0^{-i})}\left[F^i(q^i,s_0,a_0^i,a_0^{-i},s_1)\right]
\end{align*}
for all $q^i\in\mathbb{R}^{|\mathcal{S}||\mathcal{A}^i|}$.
Then Equation \ref{sa:reformulation} can be viewed as a stochastic approximation algorithm for solving the (time-varying) equation $\Bar{F}_k^i(q^i)=0$ with time-inhomogeneous Markovian noise $\{(S_k,A_k^i,A_k^{-i},S_{k+1})\}_{k\geq 0}$. We next establish the properties of the operators $F^i(\cdot)$ and $\Bar{F}_k^i(\cdot)$ in the following lemma.

\begin{lemma}[Lemma A.9 of \cite{chen2023finite}]\label{lem:operators}
	The following inequalities hold:
	\begin{enumerate}
		\item for all $(q_1^i,q_2^i)$ and $(s_0,a_0^i,a_0^{-i},s_1)$: $$\|F^i(q_1^i,s_0,a_0^i,a_0^{-i},s_1)-F^i(q_2^i,s_0,a_0^i,a_0^{-i},s_1)\|_2\leq \|q_1^i-q_2^i\|_2.$$
		\item for all $ (s_0,a_0^i,a_0^{-i},s_1)$: $\|F^i(\bm{0},s_0,a_0^i,a_0^{-i},s_1)\|_2\leq \frac{1}{1-\gamma}$.
		\item $\bar{F}_k^i(q^i)=0$ has a unique solution $\bar{q}_k^i$, which is explicitly given as $\bar{q}_k^i(s)=\mathcal{T}^i(v^i)(s)\pi_k^{-i}(s)$ for all $s$.
		\item $\langle \Bar{F}_k^i(q_1^i)-\Bar{F}_k^i(q_2^i),q_1^i-q_2^i\rangle\leq   -c_\eta\|q_1^i-q_2^i\|_2^2$ for all $(q_1^i,q_2^i)$.
	\end{enumerate}
\end{lemma}

Using $\|\cdot\|_2^2$ as a Lyapunov function and the equivalent update equation (\ref{sa:reformulation}), we obtain
\begin{align}
	\mathbb{E}[\|q_{k+1}^i-\bar{q}_{k+1}^i\|_2^2]	=\;&\mathbb{E}[\|q_{k+1}^i-q_k^i+q_k^i-\bar{q}_k^i+\bar{q}_k^i-\bar{q}_{k+1}^i\|_2^2]\nonumber\\
	=\;&\mathbb{E}[\|q_k^i-\bar{q}_k^i\|_2^2]+\mathbb{E}[\|q_{k+1}^i-q_k^i\|_2^2]+\mathbb{E}[\|\bar{q}_k^i-\bar{q}_{k+1}^i\|_2^2]\nonumber\\
	&+\alpha_k\mathbb{E}[\langle F^i(q_k^i,S_k,A_k^i,A_k^{-i},S_{k+1}),q_k^i-\bar{q}_k^i\rangle]+\mathbb{E}[\langle q_{k+1}^i-q_k^i,\bar{q}_k^i-\bar{q}_{k+1}^i\rangle]\nonumber\\
	&+\mathbb{E}[\langle q_k^i-\bar{q}_k^i,\bar{q}_k^i-\bar{q}_{k+1}^i\rangle]\nonumber\\
	=\;&\mathbb{E}[\|q_k^i-\bar{q}_k^i\|_2^2]+\alpha_k\underbrace{\mathbb{E}[\langle \bar{F}_k^i(q_k^i),q_k^i-\bar{q}_k^i\rangle]}_{N_1}\nonumber\\
	&+\alpha_k\underbrace{\mathbb{E}[\langle F^i(q_k^i,S_k,A_k^i,A_k^{-i},S_{k+1})-\bar{F}_k^i(q_k^i),q_k^i-\bar{q}_k^i\rangle]}_{N_2}\nonumber\\
	&+\mathbb{E}[\|q_{k+1}^i-q_k^i\|_2^2]+\mathbb{E}[\|\bar{q}_k^i-\bar{q}_{k+1}^i\|_2^2]\nonumber\\
	&+\mathbb{E}[\langle q_{k+1}^i-q_k^i,\bar{q}_k^i-\bar{q}_{k+1}^i\rangle]+\mathbb{E}[\langle q_k^i-\bar{q}_k^i,\bar{q}_k^i-\bar{q}_{k+1}^i\rangle].\label{eq:all_terms}
\end{align}
What remains to do is to bound the terms on the r.h.s of the previous inequality. Among them, we want to highlight the two terms $N_1$ and $N_2$. For the term $N_1$, using Lemma \ref{lem:operators} (4), we have
\begin{align}\label{eq:q_drift}
	N_1=\mathbb{E}[\langle \bar{F}_k^i(q_k^i),q_k^i-\bar{q}_k^i\rangle]
	=\mathbb{E}[\langle \bar{F}_k^i(q_k^i)-\bar{F}_k^i(\bar{q}_k^i),q_k^i-\bar{q}_k^i\rangle]
	\leq -c_\eta\mathbb{E}[\|q_k^i-\bar{q}_k^i\|_2^2],
\end{align}
which provides us with the desired drift inequality.

The term $N_2$ is the difference between $F^i(q_k^i,S_k,A_k^i,A_k^{-i},S_{k+1})$ and its expected value $\bar{F}_k^i(q_k^i)$. To analyze the latter, under the time-inhomogeneous Markov chain $\{(S_k,A_k^i,A_k^{-i},S_{k+1})\}$, we note that the strategy is updated slower than the $q$-functions; and the stationary distribution is Lipschitz in the underlying strategy (Lemma \ref{lem:zhang}). This reasoning is at the core of the proof of this result.

\begin{lemma}[Noise term, Lemma A.12 of \cite{chen2023finite}]
\label{lem:noise}
	When $\alpha_{k-\tau_k,k-1}\leq 1/4$ for all $k\geq \tau_k$, we have for all $k\geq \tau_k$ that
	\begin{align*}
		N_2\leq \frac{340|\mathcal{S}|^{3/2} A_{\max}^{3/2}L_\eta}{(1-\gamma)^2}\tau_k\alpha_{k-\tau_k,k-1}.
	\end{align*}
\end{lemma}

For our choice of $\alpha_k$ and $\beta_k$, we find $\lim_{k\rightarrow \infty}\tau_k \alpha_{k-\tau_k,k-1}=0$. Therefore, Lemma \ref{lem:noise} implies that the term $N_2$ vanishes as a function of the episode length.

We next bound the rest of the terms on the r.h.s of Equation \ref{eq:all_terms} in the following lemma.

\begin{lemma}[Other terms]\label{lem:other_terms}
	The following inequalities
	hold for all $k\geq 0$.
	\begin{enumerate}
		\item $\mathbb{E}[\|q_{k+1}^i-q_k^i\|_2^2]\leq \frac{4|\mathcal{S}| A_{\max}\alpha_k^2}{(1-\gamma)^2}$.
		\item $\mathbb{E}[\|\bar{q}_k^i-\bar{q}_{k+1}^i\|_2^2]\leq \frac{4|\mathcal{S}| A_{\max}\beta_k^2}{(1-\gamma)^2}$.
		\item $\mathbb{E}[\langle q_{k+1}^i-q_k^i,\bar{q}_k^i-\bar{q}_{k+1}^i\rangle]\leq \frac{4|\mathcal{S}| A_{\max}\alpha_k\beta_k}{(1-\gamma)^2}$.
		\item $
		\mathbb{E}[\langle q_k^i-\bar{q}_k^i,\bar{q}_k^i-\bar{q}_{k+1}^i\rangle]
		\leq \frac{17 \eta  A_{\max}^2\beta_k}{(1-\gamma)^2}\mathbb{E}[\| q_k^i-\bar{q}_k^i\|_2^2]+\frac{\beta_k}{16}\sum_{s}\mathbb{E}[V_{v,s}(\pi_k^i(s),\pi_k^{-i}(s))]$.
	\end{enumerate}
\end{lemma}

\begin{proof}
    This lemma was proved in \citet{chen2023finite}, we only provide the proof of the fourth part as it requires an adaptation due to using Tsallis entropy.

    \textit{4)} For any $k\geq 0$, we have
	\begin{align}
		\langle q_k^i-\bar{q}_k^i,\bar{q}_k^i-\bar{q}_{k+1}^i\rangle = \;& \beta_k\sum_{s}\langle q_k^i(s)-\bar{q}_k^i(s),\mathcal{T}^i(v^i)(s)(\ts(q_k^{-i}(s))-\pi_k^{-i}(s))\rangle\nonumber\\
		\leq \;&\beta_k\left(\frac{c\| q_k^i-\bar{q}_k^i\|_2^2}{2}+\frac{\sum_{s}\|\mathcal{T}^i(v^i)(s)(\ts(q_k^{-i}(s))-\pi_k^{-i}(s))\|_2^2}{2c}\right),\label{eq:otherterms_2}
	\end{align}
	where $c$ is an arbitrary positive real number.
	We next analyze the second term on the r.h.s of the previous inequality. For any $s\in\mathcal{S}$, we have
	\begin{align*}
    	\|\mathcal{T}^i(&v^i)(s)(\ts(q_k^{-i}(s))-\pi_k^{-i}(s))\|_2 \\
        = &\|\mathcal{T}^i(v^i)(s)(\ts(q_k^{-i}(s))-\ts(\bar{q}_k^{-i}(s))+\ts(\mathcal{T}^{-i}(v^{-i})(s)\pi_k^i(s))-\pi_k^{-i}(s))\|_2\\
    		\leq \;&\underbrace{\|\mathcal{T}^i(v^i)(s)(\ts(q_k^{-i}(s))-\ts(\bar{q}_k^{-i}(s)))\|_2}_{B_1} + \underbrace{\|\mathcal{T}^i(v^i)(s)(\ts(\mathcal{T}^{-i}(v^{-i})(s)\pi_k^i(s))-\pi_k^{-i}(s))\|_2}_{B_2}.
	\end{align*}
	Since Tsallis--smoothing $\ts(\cdot)$ is $2\sqrt{2} \eta  A_{\max}$ -- Lipschitz continuous with respect to $\|\cdot\|_2$, we have
	\begin{align*}
		B_1\leq \;&\|\mathcal{T}^i(v^i)(s)\|_2\|\ts(q_k^{-i}(s))-\ts(\bar{q}_k^{-i}(s))\|_2\\
		\leq \;&\frac{2\sqrt{2} \eta  A_{\max}^2}{1-\gamma}\|q_k^{-i}(s)-\bar{q}_k^{-i}(s)\|_2.
	\end{align*}
	We next analyze the term $B_2$. Using the quadratic growth property of strongly convex functions, we obtain
	\begin{align*}
		B_2=\;&\|\mathcal{T}^i(v^i)(s)(\ts(\mathcal{T}^{-i}(v^{-i})(s)\pi_k^i(s))-\pi_k^{-i}(s))\|_2\\
		\leq \;&\|\mathcal{T}^i(v^i)(s)\|_2\|\ts(\mathcal{T}^{-i}(v^{-i})(s)\pi_k^i(s))-\pi_k^{-i}(s)\|_2\\
		\leq \;&\frac{\sqrt{2\eta } A_{\max}}{1-\gamma}V_{v,s}^{1/2}(\pi_k^i(s),\pi_k^{-i}(s)).
	\end{align*}
	Combining the upper bounds we obtained for the terms $B_1$ and $B_2$ 
	\begin{align*}
		\sum_{s}\|\mathcal{T}^i(v^i)(s)&(\ts(q_k^{-i}(s))-\pi_k^{-i}(s))\|_2^2 \\
        \leq \;&\sum_{s}(B_1+B_2)^2\\
		\leq \;&2\sum_{s}(B_1^2+B_2^2)\\
		\leq \;&2\sum_{s}\left(\frac{8  A_{\max}^4 \eta^2}{(1-\gamma)^2}\|q_k^{-i}(s))-\bar{q}_k^{-i}(s)\|_2^2+\frac{2\eta  A_{\max}^2}{(1-\gamma)^2}V_{v,s}(\pi_k^i(s),\pi_k^{-i}(s))\right)\\
		= \;&\frac{16 \eta^2  A_{\max}^4}{(1-\gamma)^2}\|q_k^{-i}-\bar{q}_k^{-i}\|_2^2+\frac{4 \eta  A_{\max}^2}{(1-\gamma)^2}\sum_{s}V_{v,s}(\pi_k^i(s),\pi_k^{-i}(s)).
	\end{align*}
	Coming back to Equation \ref{eq:otherterms_2} and using the previous inequality
	\begin{align*}
		\langle q_k^i &-\bar{q}_k^i,\bar{q}_k^i-\bar{q}_{k+1}^i\rangle \\
        \leq \;& \beta_k\left(\frac{c\| q_k^i-\bar{q}_k^i\|_2^2}{2}+\frac{\sum_{s}\|\mathcal{T}^i(v^i)(s)(\ts(q_k^{-i}(s))-\pi_k^{-i}(s))\|_2^2}{2c}\right)\\
		\leq \;&\beta_k\left(\frac{c\| q_k^i-\bar{q}_k^i\|_2^2}{2}+ \frac{8 \eta^2  A_{\max}^4}{c (1-\gamma)^2}\|q_k^{-i}-\bar{q}_k^{-i}\|_2^2 +\frac{2 \eta  A_{\max}^2}{c (1-\gamma)^2}\sum_{s}V_{v,s}(\pi_k^i(s),\pi_k^{-i}(s))\right).
	\end{align*}
	Choosing $c=\frac{32 \eta  A_{\max}^2}{(1-\gamma)^2}$ in the previous inequality and then taking total expectation
	\begin{align*}
		\mathbb{E}[\langle q_k^i-\bar{q}_k^i,\bar{q}_k^i-\bar{q}_{k+1}^i\rangle]
		\leq \frac{17 \eta  A_{\max}^2\beta_k}{(1-\gamma)^2}\mathbb{E}[\| q_k^i-\bar{q}_k^i\|_2^2]+\frac{\beta_k}{16}\sum_{s}\mathbb{E}[V_{v,s}(\pi_k^i(s),\pi_k^{-i}(s))].
	\end{align*} 
\end{proof}

Since we obtained upper bounds on all the terms on the r.h.s of Equation \ref{eq:all_terms}, we deduce the one-step Lyapunov drift inequality for $q_k^i$. The same analysis also entails a drift inequality for $q_k^{-i}$. Both results are presented in the following lemma. 

\begin{lemma}[Lyapunov drift for $q$--functions, Lemma A.12 of \citet{chen2023finite}]
\label{lem:q-function-drift}
	For all $k\geq \tau_k$ and $i\in \{1,2\}$:
	\begin{align*}
		\mathbb{E}[\|q_{k+1}^i-\bar{q}_{k+1}^i\|_2^2]\leq \;&\left(1-c_\eta\alpha_k+\frac{17 \eta  A_{\max}^2\beta_k}{(1-\gamma)^2}\right)\mathbb{E}[\|q_k^i-\bar{q}_k^i\|_2^2]\\
		&+\frac{352|\mathcal{S}|^{3/2} A_{\max}^{3/2}L_\eta}{(1-\gamma)^2}\tau_k\alpha_k\alpha_{k-\tau_k,k-1}+\frac{\beta_k}{16}\sum_{s}\mathbb{E}[V_{v,s}(\pi_k^i(s),\pi_k^{-i}(s))].
	\end{align*}
 where $\tau_k=t_{\ell_\eta,\beta_k}$ is a uniform upper bound on the uniform mixing time with accuracy $\beta_k$, see Equation \ref{eq:mixing_time_definition}).
\end{lemma}

\subsection{Bounding Coupled Drift Inequalities}\label{sec:strategy}

We first restate the drift inequalities from previous sections. Recall the notations: $\mathcal{D}_q(t,k)=\sum_{i=1,2}\|q_{t,k}^i-\bar{q}_{t,k}^i\|_2^2$, $\mathcal{D}_\pi(t,k)=\sum_{s}V_{v_t,s}(\pi_{t,k}^i(s),\pi_{t,k}^{-i}(s))$, and $\mathcal{F}_t$ as the history of the algorithm right before the $t$-th outer-loop iteration. Note that $v_t^i$ and $v_t^{-i}$ are both measurable with respect to $\mathcal{F}_t$. In what follows, we denote $\mathbb{E}_t[\;\cdot\;]$ for $\mathbb{E}[\;\cdot\;\mid \mathcal{F}_t]$. 

\begin{itemize}
	\item \textbf{Lemma \ref{lem:outer-loop}:} It holds for all $t\geq 0$ and $i=1,2$ that
	\begin{align}\label{eq:result:le:outer-loop}
		\|v_{t+1}^i-v_*^i\|_\infty
		\leq \;&\gamma  \|v_t^i-v_*^i\|_\infty+2\max_{s\in\mathcal{S}}V_{v_t,s}(\pi_{t,K}^i(s),\pi_{t,K}^{-i}(s)) + \frac{16}{\eta} \sqrt{ A_{\max}}\nonumber\\
        &+ \max_{s\in\mathcal{S}} \|\Bar{q}_{t,K}^i(s) - q_{t,K}^i(s) \|_\infty + 2 \gamma \|v_t^i+v_t^{-i}\|_\infty,
	\end{align}
        where $\Bar{q}_{t,K}^i(s):=\mathcal{T}^i(v_t^i)(s)\pi_{t,K}^{-i}(s)$ for all $s\in\mathcal{S}$.
	\item \textbf{Lemma \ref{lem:outer-sum}:} It holds for all $t\geq 0$ that
	\begin{align}\label{eq:Lyapunov_v+}
		\|v_{t+1}^i+v_{t+1}^{-i}\|_\infty\leq\;& \gamma\|v_t^i+v_t^{-i}\|_\infty+\sum_{i=1,2}\|q_{t,K}^i-\bar{q}_{t,K}^i\|_2.
	\end{align}
	\item \textbf{Lemma \ref{lem:policy_drift}: } It holds for all $t,k\geq 0$ that
	\begin{align}\label{eq:Lyapunov_pi}
		\mathbb{E}_t[\mathcal{D}_\pi(t,k+1)]
		\leq &\left(1-\frac{3\beta_k}{4}\right)\mathbb{E}_t[\mathcal{D}_\pi(t,k)]+\frac{2048  A_{\max}^4 \beta_k \eta^3}{\ell_{\eta}^3 (1-\gamma)^2}\mathbb{E}_t[\mathcal{D}_q(t,k)]\nonumber\\
		&+16|\mathcal{S}| A_{\max} \beta_k\eta \|v_t^i+v_t^{-i}\|_\infty^2+\frac{4|\mathcal{S}| A_{\max}^2\beta_k^2}{\ell_{\eta}(1-\gamma)^2}.
	\end{align}
	\item \textbf{Lemma \ref{lem:q-function-drift}:} It holds for all $t\geq 0$ and $k\geq \tau_k$ that
	\begin{align}\label{eq:Lyapunov_q}
            \mathbb{E}_t[ \mathcal{D}_q(t,k+1) ] \leq \;&\left(1-c_\eta\alpha_k+\frac{17 \eta  A_{\max}^2 \beta_k}{(1-\gamma)^2}\right)\mathbb{E}_t[\mathcal{D}_q(t,k)]\nonumber\\
		&+\frac{\beta_k}{16}\mathbb{E}_t[\mathcal{D}_\pi(t,k)]+\frac{352|\mathcal{S}|^{3/2} A_{\max}^{3/2}L_\eta}{(1-\gamma)^2}\tau_k\alpha_k\alpha_{k-\tau_k,k-1}.
	\end{align}
\end{itemize}
Adding up equations \eqref{eq:Lyapunov_pi} and \eqref{eq:Lyapunov_q} entails
\begin{align*}
    \mathbb{E}_t[ \mathcal{D}_\pi(t,k+1)& + \mathcal{D}_q(t,k+1) ] \leq \left(1-\frac{\beta_k}{2}\right)\mathbb{E}_t[\mathcal{D}_\pi(t,k)] +\frac{4|\mathcal{S}| A_{\max}^2\beta_k^2}{\ell_{\eta}(1-\gamma)^2}\\
    &+ \left(1-c_\eta\alpha_k+\frac{3136  A_{\max}^4 \beta_k \eta^3}{\ell_{\eta}^3 (1-\gamma)^2}\right)\mathbb{E}_t[\mathcal{D}_q(t,k)]+ 16|\mathcal{S}| A_{\max} \beta_k \eta  \|v_t^i+v_t^{-i}\|_\infty^2 \\
 & \quad +\frac{352|\mathcal{S}|^{3/2} A_{\max}^{3/2}L_\eta}{(1-\gamma)^2}\tau_k\alpha_k\alpha_{k-\tau_k,k-1}\\
	= \;&\left(1-\frac{\beta\alpha_k}{2\alpha}\right)\mathbb{E}_t[\mathcal{D}_\pi(t,k)]+\left(1-c_\eta\alpha_k+\frac{3136  A_{\max}^4 \beta\alpha_k \eta^3}{\alpha\ell_{\eta}^3 (1-\gamma)^2}\right)\mathbb{E}_t[\mathcal{D}_q(t,k)]\nonumber\\
	& + 16|\mathcal{S}| A_{\max} \beta_k \eta \|v_t^i+v_t^{-i}\|_\infty^2+\frac{4|\mathcal{S}| A_{\max}^2\beta_k^2}{\ell_{\eta}(1-\gamma)^2}\\
    & \quad +\frac{352|\mathcal{S}|^{3/2} A_{\max}^{3/2}L_\eta}{(1-\gamma)^2}\tau_k\alpha_k\alpha_{k-\tau_k,k-1}.
\end{align*}
Note that requirements on stepsizes in Appendix \ref{ap:sample_complexity_analysis} imply that
\begin{align*}
    \frac{3136  A_{\max}^4 \beta \eta^3}{\alpha\ell_{\eta}^3 (1-\gamma)^2}\leq \frac{c_\eta}{2}.
\end{align*}
Therefore, we have
\begin{align}\label{eq:before_stepsize}
	\mathbb{E}_t[ \mathcal{D}_\pi(t,k+1) + \mathcal{D}_q(t,k+1)] \le &\frac{ 4|\mathcal{S}|  A_{\max}^2 \beta^2\alpha_k^2 }{\alpha^2\ell_{\eta}(1-\gamma)^2}+\frac{352|\mathcal{S}|^{3/2} A_{\max}^{3/2}L_\eta}{(1-\gamma)^2}\tau_k\alpha_k\alpha_{k-\tau_k,k-1} \nonumber\\
	&+  \left(1-\frac{\beta\alpha_k}{2\alpha}\right)\mathbb{E}_t[\mathcal{D}_\pi(t,k)+\mathcal{D}_q(t,k)] \\
    & + 16|\mathcal{S}| A_{\max} \frac{\beta}{\alpha}\alpha_k \eta \|v_t^i+v_t^{-i}\|_\infty^2.
\end{align}

Recall that Theorem \ref{thm:NashGapBound} recommends choosing $\alpha_k=\frac{\alpha}{k+h}$, $\beta_k=\frac{\beta}{k+h}$, with certain requirements for $\alpha$ and $\beta$. Iteratively applying Equation \ref{eq:before_stepsize}, we obtain for all $k\geq k_0$
\begin{align*}
	\mathbb{E}_t[\mathcal{D}_\pi(t,k)+\mathcal{D}_q(t,k)]
	\lesssim &\frac{4|\mathcal{S}| A_{\max}}{(1-\gamma)^2}\underbrace{\prod_{m=k_0}^{k-1}\left(1-\frac{\beta\alpha_m}{2\alpha}\right)}_{\hat{\mathcal{E}}_1}\nonumber\\
	&+\frac{|\mathcal{S}|^{3/2} A_{\max}^2L_\eta}{(1-\gamma)^2}\underbrace{\sum_{n=k_0}^{k-1}z_n^2\alpha_n^2\prod_{m=n+1}^{k-1}\left(1-\frac{\beta\alpha_m}{2\alpha}\right)}_{\hat{\mathcal{E}}_2}\\
	&+ |\mathcal{S}| A_{\max} \frac{\beta}{\alpha} \eta \|v_t^i+v_t^{-i}\|_\infty^2\underbrace{\sum_{n=k_0}^{k-1}\alpha_n\prod_{m=n+1}^{k-1}\left(1-\frac{\beta\alpha_m}{2\alpha}\right)}_{\hat{\mathcal{E}}_3}.
\end{align*}
We next provide estimates for the terms $\{\hat{\mathcal{E}}_j\}_{1\leq j\leq 3}$. Bounds of terms like $\{\hat{\mathcal{E}}_j\}_{1\leq j\leq 3}$ are well-established in existing work studying the convergence rate of iterative algorithms. Specifically, we have that
\begin{align*}
	\hat{\mathcal{E}}_1\leq
	\left(\frac{k_0+h}{k+h}\right)^{\beta/2},\quad 
	\hat{\mathcal{E}}_2\leq
	\frac{4e\tau_k^2\alpha^2}{\beta/2-1}\frac{1}{k+h},\;\text{ and }\;
	\hat{\mathcal{E}}_3\leq \frac{2\alpha}{\beta}.
\end{align*}
It follows that 
\begin{align}\label{eq:ci_di}
	\mathbb{E}_t[\mathcal{D}_\pi(t,k)\!+\!\mathcal{D}_q(t,k)] \lesssim &\frac{|\mathcal{S}| A_{\max}}{(1-\gamma)^2}\!\left(\frac{k_0+h}{k+h}\right)^{\beta/2} \!\!\!\!+\! \frac{|\mathcal{S}|^{3/2} A_{\max}^2L_\eta}{(1-\gamma)^2}\frac{\tau_k^2\alpha^2}{\beta/2-1}\frac{1}{k+h} \!+\! |\mathcal{S}| A_{\max} 
 \eta \|v_t^i+v_t^{-i}\|_\infty^2\nonumber\\
	\lesssim  \;&\frac{|\mathcal{S}| A_{\max}}{(1-\gamma)^2}\left(\frac{\alpha_k}{\alpha_{k_0}}\right)^{\beta/2}+\frac{|\mathcal{S}|^{3/2} A_{\max}^2L_\eta}{(1-\gamma)^2}\frac{\tau_k^2\alpha^2}{\beta/2-1}\frac{1}{k+h} + |\mathcal{S}| A_{\max} 
 \eta \|v_t^i+v_t^{-i}\|_\infty^2,
\end{align}
which implies
\begin{align*}
	\mathbb{E}_t[\mathcal{D}_\pi(t,k)]
	\lesssim  \;&\frac{|\mathcal{S}| A_{\max}}{(1-\gamma)^2}\left(\frac{\alpha_k}{\alpha_{k_0}}\right)^{\beta/2}+\frac{|\mathcal{S}|^{3/2} A_{\max}^2L_\eta}{(1-\gamma)^2}\frac{\tau_k^2\alpha^2}{\beta/2-1}\frac{1}{k+h} + |\mathcal{S}| A_{\max} \eta \|v_t^i+v_t^{-i}\|_\infty^2.
\end{align*}
Using the previous bound on $\mathbb{E}_t[\mathcal{D}_\pi(t,k)]$ in Equation \ref{eq:Lyapunov_q}
\begin{align*}
	\mathbb{E}_t[\mathcal{D}_q(t,k+1)]\leq \;&\left(1-c_\eta\alpha_k+\frac{17 \eta  A_{\max}^2\beta_k}{(1-\gamma)^2}\right)\mathbb{E}_t[\mathcal{D}_q(t,k)]+\frac{\beta_k}{16}\mathbb{E}_t[\mathcal{D}_\pi(t,k)]\\
    &+\frac{352|\mathcal{S}|^{3/2} A_{\max}^{3/2}L_\eta}{(1-\gamma)^2}\tau_k\alpha_k\alpha_{k-\tau_k,k-1}\\
	\lesssim \;&\left(1-\frac{c_\eta\alpha_k}{2}\right)\mathbb{E}_t[\mathcal{D}_q(t,k)]+\frac{|\mathcal{S}|^{3/2} A_{\max}^2}{\alpha_{k_0}(1-\gamma)^2}\tau_k^2\alpha_k^2 + |\mathcal{S}| A_{\max} \frac{\beta}{\alpha}\alpha_k \eta \|v_t^i+v_t^{-i}\|_\infty^2.
\end{align*}
Repeatedly using the previous inequality starting from $k_0$ 
\begin{align*}
	\mathbb{E}_t[\mathcal{D}_q(t,k)]\lesssim\;&\frac{|\mathcal{S}| A_{\max}}{(1-\gamma)^2}\left(\frac{\alpha_k}{\alpha_{k_0}}\right)^{c_\eta\alpha/2}+\frac{|\mathcal{S}|^{3/2} A_{\max}^2}{\alpha_{k_0}(1-\gamma)^2}\tau_k^2\alpha_k +\frac{|\mathcal{S}| A_{\max} \beta \eta }{\alpha c_\eta}\|v_t^i+v_t^{-i}\|_\infty^2\\
	\lesssim\;&\frac{|\mathcal{S}|^{3/2} A_{\max}^2}{\alpha_{k_0}(1-\gamma)^2}\tau_k^2\alpha_k+\frac{|\mathcal{S}| A_{\max} \beta \eta }{\alpha c_\eta}\|v_t^i+v_t^{-i}\|_\infty^2
\end{align*}
Since $\sum_{i=1,2}\mathbb{E}_t\left[\|q_{t,K}^i-\bar{q}_{t,K}^i\|_2\right]\lesssim \mathbb{E}_t[\mathcal{D}_q(t,K)]^{1/2}$, 
we have
\begin{align}\label{eeeq}
	\sum_{i=1,2}\mathbb{E}_t\left[\|q_{t,K}^i-\bar{q}_{t,K}^i\|_2\right]\leq\;&\frac{c_1'|\mathcal{S}|^{3/4} A_{\max}}{\alpha^{1/2}_{k_0}(1-\gamma)}\tau_k\alpha_k^{1/2}+\frac{c_2'\sqrt{|\mathcal{S}| A_{\max}}c^{1/2}_{\alpha,\beta} \eta^{1/2}}{c_{\eta}^{1/2}}\|v_t^i+v_t^{-i}\|_\infty,
\end{align}
where $c_1'$ and $c_2'$ are numerical constants.
Applying the total expectation for both sides of the previous inequality and using Equation \ref{eq:Lyapunov_v+}, we obtain
\begin{align*}
	\mathbb{E}[\|v_{t+1}^i+v_{t+1}^{-i}\|_\infty] \leq\;& \left(\gamma+\frac{c_2'\sqrt{|\mathcal{S}| A_{\max}}c^{1/2}_{\alpha,\beta} \eta^{1/2}}{c_{\eta}^{1/2}}\right)\mathbb{E}[\|v_t^i+v_t^{-i}\|_\infty]+\frac{c_1'|\mathcal{S}|^{3/4} A_{\max}}{\alpha^{1/2}_{k_0}(1-\gamma)}\tau_k\alpha_k^{1/2}\\
	\leq \;&\left(\frac{\gamma+1}{2}\right)\mathbb{E}[\|v_t^i+v_t^{-i}\|_\infty]+\frac{c_1'|\mathcal{S}|^{3/4} A_{\max}}{\alpha^{1/2}_{k_0}(1-\gamma)}\tau_k\alpha_k^{1/2},
\end{align*}
where the last line follows from the requirements on stepsizes in Appendix \ref{ap:sample_complexity_analysis}. Repeatedly using the previous inequality starting from $0$
\begin{align}\label{eq:zls}
	\mathbb{E}[\|v_t^i+v_t^{-i}\|_\infty]\lesssim \frac{2}{1-\gamma}\left(\frac{\gamma+1}{2}\right)^t+\frac{|\mathcal{S}|^{3/4} A_{\max}}{\alpha^{1/2}_{k_0}(1-\gamma)^2}\tau_k\alpha_k^{1/2}.
\end{align}
The next step is to bound $\|v_t^i-v_*^i\|_\infty$. Recall from Equation \ref{eq:result:le:outer-loop} that
\begin{align*}
    \mathbb{E}[\|v_{t+1}^i-v_*^i\|_\infty]
	\leq \;\gamma  \mathbb{E}[\|v_t^i-v_*^i\|_\infty]+2 \gamma \mathbb{E}[\|v_t^i+v_t^{-i}\|_\infty]+\frac{16}{\eta} \sqrt{ A_{\max}}+2\mathbb{E}[\mathcal{D}_\pi(t,K)]+2\mathbb{E}[\mathcal{D}_q(t,K)]^{1/2}.
\end{align*}
Since equations \ref{eq:ci_di} and \ref{eeeq} imply that
\begin{align*}
    \mathbb{E}[\|v_t^i+v_t^{-i}\|_\infty] + \mathbb{E}[\mathcal{D}_\pi(t,K)]+\mathbb{E}[\mathcal{D}_q(t,K)]^{1/2}
	\lesssim \;\frac{|\mathcal{S}|A_{\max} \eta}{(1-\gamma)^2}\left(\frac{\gamma+1}{2}\right)^t+\frac{|\mathcal{S}|^2A_{\max}^2L_\eta \alpha}{\alpha_{k_0} \beta(1-\gamma)^3}\tau_K^2\alpha_K^{1/2},
\end{align*}
we have 
\begin{align*}
    \mathbb{E}[\|v_{t+1}^i-v_*^i\|_\infty]\leq\;& \gamma  \mathbb{E}[\|v_t^i-v_*^i\|_\infty]+c''\Bigg[\frac{|\mathcal{S}| \eta A_{\max} }{(1-\gamma)^2}\left(\frac{\gamma+1}{2}\right)^t+\frac{ \sqrt{A_{\max}}}{\eta} + \frac{|\mathcal{S}|^2A_{\max}^2L_\eta \alpha}{\alpha_{k_0} \beta(1-\gamma)^3}\tau_K^2\alpha_K^{1/2}\Bigg]
\end{align*}
for some numerical constant $c''$. We use the previous inequality iteratively starting from $0$ to time $T-1$ to find
\begin{align*}
	\mathbb{E}[\|v_T^i-v_*^i\|_\infty]\lesssim \frac{|\mathcal{S}|A_{\max} T \eta }{(1-\gamma)^2}\left(\frac{\gamma+1}{2}\right)^{T-1}+\frac{ \sqrt{A_{\max}}}{\eta (1-\gamma)}+\frac{|\mathcal{S}|^2A_{\max}^2L_\eta \alpha}{\alpha_{k_0} \beta(1-\gamma)^4}\tau_K^2\alpha_K^{1/2}
\end{align*}
Plugging the previous inequality with equations \ref{eq:ci_di} and \ref{eq:zls} in the Nash gap decomposition (lemma \ref{lem:Nash_Gap_Decomp}), we obtain
\begin{align*}
	\mathbb{E}[\|v^i_{*,\pi_{T,K}^{-i}}-v^i_{\pi_{T,K}^i,\pi_{T,K}^{-i}}\|_\infty] \lesssim \;&\frac{|\mathcal{S}|A_{\max} T \eta }{ (1-\gamma)^3}\left(\frac{\gamma+1}{2}\right)^{T-1}+\frac{ \sqrt{A_{\max}}}{\eta (1-\gamma)^2} + \frac{|\mathcal{S}|^2A_{\max}^2L_\eta\alpha}{\alpha_{k_0} \beta(1-\gamma)^5}\tau_K^2\alpha_K^{1/2}
\end{align*}
Finally,
\begin{align*}
	\mathbb{E}[\textit{NG}(\pi_{T,K}^i,\pi_{T,K}^{-i})] \lesssim \;&\frac{|\mathcal{S}|A_{\max} T \eta }{(1-\gamma)^3}\left(\frac{\gamma+1}{2}\right)^{T-1}+\frac{ \sqrt{A_{\max}}}{\eta (1-\gamma)^2} + \frac{|\mathcal{S}|^2A_{\max}^2L_\eta\alpha}{\alpha_{k_0} \beta(1-\gamma)^5}\tau_K^2\alpha_K^{1/2}.
\end{align*}
This concludes the proof of theorem \ref{thm:NashGapBound}.

\section{Proof of corollaries}

\subsection{Proof of corollary \ref{cor:SampleComplexity}}
\label{app:proof_of_cor_1}

We have from theorem \ref{thm:NashGapBound} that under self-play and Assumption \ref{as:diameter}, if $\frac{\beta}{\alpha} \leq \frac{c_\eta\ell_{\eta}^3 (1-\gamma)^2}{6272 \eta^3 |\mathcal{S}|A_{\max}^4}$, then Algorithm \ref{algorithm:TBRVI} achieves for all $K\geq k_0$:
	\begin{align*}
		\mathbb{E}[\textit{NG}(\pi_{T,K}^i,\pi_{T,K}^{-i})] \leq &  \frac{ c_1 |\mathcal{S}| A_{\max} T \eta }{ (1-\gamma)^3 }\left(\frac{ \gamma+1 }{2}\right)^{T-1} + \frac{ c_2|\mathcal{S}|^2A_{\max}^2 L_\eta \alpha}{\alpha_{k_0} \beta(1-\gamma)^5}\frac{\tau_K^2\alpha^{1/2}}{(K+h)^{1/2}} +\frac{c_3 \sqrt{A_{\max}}}{\eta (1-\gamma)^2}.
	\end{align*}

In the following, we will provide bounds on the terms that appear in the bound above and that depend on $\eta$. Namely, we show that:

\begin{align*}
    \tau_K &= \mathcal{O}(\log(K)/(\ell_\eta^{2 d_r} \mu_{r,min})), \\
    L_\eta &= \mathcal{O}(\ell_\eta^{-2 d_r})
\end{align*}

First, we prove these inequalities, then we use them to deduce the sample complexity.

\textbf{Proof of inequalities} 

\textbf{\textit{1)}} We know that $\tau_K = t_{\ell_\eta,\beta_k} \le \frac{t_{\pi_r,\beta_k}}{\ell_\eta^{2 d_r} \mu_{r,min}}$ and $t_{\pi_r,\beta_K} = \mathcal{O}(\log (1/\beta_K)$ thanks to the fast mixing of $\pi_r$. Therefore, with the choice $\beta_K \propto 1/K$ we obtain that
 \begin{equation*}
     \tau_K = \mathcal{O}(\log(K)/(\ell_\eta^{2 d_r} \mu_{r,min})).
 \end{equation*}

 \textbf{\textit{2)}} We have that $L_\eta:=\frac{2\log(8|\mathcal{S}|/\rho_\eta)}{\log(1/\rho_\eta)}$ with $\rho_\eta=\rho_r^{(\ell_\eta^2)^{d_r}\mu_{r,\min}}$, therefore as $\ell_\eta \to 0$ we obtain $L_\eta = \mathcal{O}(\ell_\eta^{-2 d_r})$.

 

\textbf{Combination:} Given Lemma \ref{lem:margins}, we have $\ell_\eta = \mathcal{O}(\eta^{-2})$, therefore, the condition on $\alpha$ and $\beta$ becomes
\begin{align*}
    \frac{\beta}{\alpha} &\leq \frac{c_\eta\ell_{\eta}^3 (1-\gamma)^2}{6272 \eta^3 |\mathcal{S}|A_{\max}^4} \le \frac{\mu_{r,min}\ell_\eta^3 \ell_\eta^3 (1-\gamma^2)}{6272 \eta^3 |\mathcal{S}|A_{\max}^4} = \frac{\mu_{r,min} (1-\gamma^2)}{6272 \eta^{15} |\mathcal{S}|A_{\max}^4}
\end{align*}

By injecting the obtained bounds in theorem \ref{thm:NashGapBound} we find for all $K\geq k_0$:
\begin{align*}
    \mathbb{E}[\textit{NG}(\pi_{T,K}^i,\pi_{T,K}^{-i})] = \mathcal{O} \left( \eta\left(\frac{ \gamma+1 }{2}\right)^{T-1} + \eta^{12 d_r +15}\frac{\log(K)^2}{\sqrt{K}} +\frac{1}{\eta} \right).
\end{align*}

The first term in the r.h.s above is exponentially decaying in $T$. We optimize the choice of $\eta$ for the two other terms and obtain the desired result for $\eta = K^{1/(24 d_r +32)}$.

\subsection{Proof of Corollary \ref{cor:rationality}}

In this section, we show that if the opponent were to fix their policy, then the player converges to the best response of the opponent. Our argument here is inspired by \cite{sayin2021decentralized}, who used it to prove the rationality of decentralized $Q$-learning.

First, we argue that the Nash gap bound in theorem \ref{thm:NashGapBound} remains true for noisy rewards as long as the zero-sum structure is preserved. Indeed, if we generalize the rewards for actions $(a^i , a^{-i})$ from $\mathcal{R}^i(a^i,a^{-i})$ to $r^i(a^i,a^{-i},\xi)$, where $\xi\in \Xi$ is a random variable with distribution $\mu_\xi$ over $\Xi$ (a finite set). If the noise is independent of the actions, $r^i+r^{-i}=0$, and the reward is still uniformly bounded, then the proof still holds.

Second, observe that if the opponent follows a stationary strategy $\pi^{-i}$, then it can be seen as additional randomness in the rewards. Specifically, consider a fictitious opponent with one available action $a^*$, and define $\hat{r}^i(a^i,a^*,a^{-i})=\mathcal{R}^i(a^i,a^{-i})$, $\hat{p}(s'\mid s,a^i,a^*)=\sum_{\pi^{-i}(a^{-i}|s)}P(s'\mid a^i,a^{-i},s)$ for all $(a^i ,a^{-i})$. In this new zero-sum game, we can apply Corollary \ref{cor:SampleComplexity} because our assumption of existence of $\pi_i$ ensuring an irreducible and aperiodic Markov chain translates into Assumption \ref{as:diameter} in this fictional setting. Therefore, Corollary \ref{cor:SampleComplexity} entails the same sample complexity for player $i$ to find an approximate best response of its opponent.

\section{Limitations of common assumptions: an illustrating example}
\label{app:counteexample}

Here we take a deeper look at the MDP of Figure \ref{fig:MDP_exp} with the strategy $\pi$ parameterized by $\xi \in [0,1]$ defined in Equation \ref{eq:exp_policy} as $\pi(1,a) = \xi \text{ and } \quad \pi(1,b) = 1-\xi.$




Using this strategy in the provided MDP example yields the following transition matrix:
\begin{align*}
    P_\xi \eqdef \begin{bmatrix}
        \frac{1-\xi}{2} & \frac{1}{2} & \frac{\xi}{2}\\
        \frac{1}{2} &  \frac{1}{2} & 0\\
        \frac{1}{2} & \frac{1}{2} & 0 
    \end{bmatrix}
\end{align*}

Therefore, to obtain the corresponding stationary distribution we solve the following for $\mu_\pi = (x,y,z) \in [0,1]^3$ and $x+y+z=1$:
\begin{equation*}
    (x,y,z) P_\xi = (x,y,z)
\end{equation*}
which is equivalent to
\begin{align*}
    \left\{
    \begin{array}{c}
    \left(\frac{1-\xi}{2}\right) x+\frac{1}{2} y+\frac{1}{2} z=x \\
    \frac{1}{2}(x+y+z)=y \\
    \frac{\xi}{2} x=z \\
    x+y+z = 1
    \end{array}
    \right.
\end{align*}
then, 
\begin{align*}
    \left\{
    \begin{array}{c}
    z = \frac{\xi}{2}x \\
    y = \frac{2+\xi}{2}x \\
    x(2+\xi) = 1
    \end{array}
    \right.
\end{align*}
from which we deduce that the stationary distribution is given by:
\begin{equation*}
    \mu_\pi = \left(\frac{1}{2+\xi}, \frac{1}{2}, \frac{\xi}{4+2\xi}\right).
\end{equation*}

\paragraph{Strong reachability} From the explicit formula of the stationary distribution $\mu_\pi$ we can deduce that as $\xi \to 0$ we have $\mu_\pi(3) \to 0 $. Then, using Equation \ref{eq:Equivalent_Reachability} we deduce that $T_{s\to s}^\pi \to \infty$. This entails that the strong reachability assumption does not hold in this simple setting. More generally, it is intuitive that stationary distributions of nearly deterministic policies would not have a full support in any non-trivial MDP.

\paragraph{Mixing time} Now we shift our focus to the mixing time assumption and we show that the mixing time for this simple MDP and strategy can grow arbitrarily if $\xi \to 1$. Consider an initial distribution $\mu_0 = (1/2,1/4,1/4)$, and denote by $x_k$ the probability that the state at time $k$ is the zero state. Our goal in the following is to show that the time it takes for $x_k$ to converge to its stationary value can grow arbitrarily if $\xi \to 1$.

First, using a one step transition we have that:
\begin{align*}
    x_{k+1} &= \frac{1-\xi}{2}x_k + \frac{1}{4} + \frac{1}{2} * \frac{\xi}{2} x_{k-1},
\end{align*}
which can be equivalently written as:
\begin{equation*}
    x_{k+1} - \frac{1-\xi}{2} x_k - \frac{\xi}{4} x_{k-1} - \frac{1}{4} = 0
\end{equation*}
and
\begin{equation*}
    \left(x_{k+1}-\frac{1}{2+\xi}\right) - \frac{1-\xi}{2} \left(x_k-\frac{1}{2+\xi}\right) - \frac{\xi}{4} \left(x_{k-1}-\frac{1}{2+\xi}\right) = 0.
\end{equation*}

The above is a homogeneous linear recurrence relation with constant coefficients and can be solved by solving the second-degree equation in $r$: $r^2 - \frac{1-\xi}{2} r - \frac{\xi}{4} = 0$. The two solutions of this equation are $r_{1} = \frac{1-\xi}{4}-\sqrt{\left(\frac{1-\xi}{4}\right)^2 + \xi}$ and  $r_{2} = \frac{1-\xi}{4}+\sqrt{\left(\frac{1-\xi}{4}\right)^2 + \xi}$ which entails the existence of $\alpha_\xi , \beta_\xi \in \bR$ such that:
\begin{equation*}
    x_{k} = \alpha_\xi r_1^k + \beta_\xi r_2^k + \frac{1}{2+\xi}.
\end{equation*}

Going back to the mixing time, we have that:
\begin{align*}
    t_{\pi, \epsilon} & =\min\left\{k \geq 0 \: :\max _{s\in \cS } \left\| P_\xi^k(\cdot | s) - \mu_\pi(\cdot) \right\|_{\mathrm{TV}} \leq \epsilon\right\} \\
    & \geq \min _{k \geq 0}\left\{\left\|P_\xi^{2k} (0 |s) - \frac{1}{2+\xi}\right\|_{\mathrm{TV}} \leq \epsilon \right\} \\
    & =\min _{k \geq 0}\left\{(\alpha_\xi r_1^{2k} + \beta_\xi r_2^{2k}) \leq \epsilon \right\} \\
    & \ge \min _{k \geq 0}\left\{\beta_\xi r_2^{2k} \leq \epsilon - \alpha_\xi r_1^{2k} \right\}\\
    & \geq \frac{\log (\beta_\xi / (\epsilon - \alpha_\xi r_1^{2k}))}{\log\left(\frac{1}{(1-\xi)^2 /8 + \xi}\right)} - 1 ,
\end{align*}

Using the initial distribution $\mu_0= (1/2,1/4,1/4)$ and $P_\xi$ we can easily deduce that for $\xi \to 1$ we obtain $r_1 \to -1$ and $r_2 \to 1$ and $\alpha\xi \to  1/8$ $\beta_\xi \to  1/24$. 

Consequently, we have for any $\epsilon \ge 1/3$, when $\xi \to 1$:
\[t_{\pi, \epsilon}  \approx \frac{\log \frac{1}{24 \epsilon-3}}{\log\frac{1}{\xi}} - 1 \]
which implies that:
\begin{equation*}
    t_{\pi, \epsilon} \to \infty \quad \text{ when } \xi \to 1.
\end{equation*}

The mixing time can grow arbitrarily if $\xi \to 1$, thus invalidating the uniform mixing-time assumption (see Definition \ref{as:bounded_mixing}). This completes our proof.

\begin{remark}
    This computation could also be done with other initial distributions $\mu_0$ as long as they entail non-zero limits for $\beta_\xi$ when $\xi \to 1$. Indeed, the second root $r_2$ converges to $1$ when $\xi$ tends to $1$ independently from the initial conditions, and this is the core insight behind this counterexample.
\end{remark}